# Solar Type IV bursts at frequencies 10-30 MHz

**V.N. Melnik (1), H. O. Rucker (2), A. A. Konovalenko (1), V.V. Dorovskyy (1), E.P. Abranin (1), A.I. Brazhenko (3), B.Thide (4), A. A. Stanislavskyy (1)**

  (1) Institute of Radio Astronomy, Kharkov, Ukraine,
  (2) Space Research Institute, Graz, Austria,
  (3) Poltava Gravimetrical Observatory, Poltava, Ukraine
  (4) Swedish Institute of Space Physics, Uppsala, Sweden

## Abstract

The results of the first observations of Type IV bursts at frequencies 10-30 MHz are presented. These observations were carried out at radio telescopes UTR-2 (Kharkov, Ukraine) and URAN-2 (Poltava, Ukraine) during the period 2003-2006. Detection of Type IV bursts in wide band from 10 to 30MHz with high sensitivity and time resolution allowed to study their properties in details. These bursts have fluxes 10-2000s.f.u. at maximum phase. Their durations are about 1-2 hours and even more. Some of Type IV bursts drift from high to low frequencies with drift rates about 10kHz/s. All observed Type IV bursts have fine structures in the form of sub-bursts with durations from 2s to 20s and frequency drift rates in a majority of 1-2MHz/s. In most cases, sub-bursts with negative drift rates were registered. Sometimes sub-bursts in absorption with durations 10-200s against Type IV burst background have been observed. The Type IV burst observed on July 22, 2004 had zebra structure, in which single zebra stripes had positive, negative and infinite drift rates.

## Introduction

Solar Type IV bursts were identified by Boischot [1] as continuous radiation appearing after flares connected with Coronal Mass Ejections (CME). There exist moving and stationary Type IV bursts [2]. The frequency of radio emission of moving bursts moves from high to low frequencies and the corresponding linear velocity of the source varies from 200 to 1600km/s [3]. The source velocity distribution is similar to the CME velocity distribution and Type IV bursts often follow Type II bursts caused by the shock waves. Apparently, the Type IV bursts are connected with shock waves, which are responsible for Type II bursts.

Sometimes Type II bursts are observed on the background of Type IV ones and thus are called Type II-IV bursts. The Type IV bursts are registered in the wide frequency band – from 408MHz to 18MHz [2], and more recently at lower frequencies from 14 to 1MHz [4, 5]. The duration of Type IV bursts decreases with increasing frequency: the Type IV's duration at 80MHz is in average 30min, but at much lower frequency it rises up to 2 hours [6].

The flux of a burst at fixed frequency first increases and then decreases with approximately equal rates. At the same time, the polarization degree keeps increasing during all the burst lifetime and sometimes reaches 90% [7]. The brightness temperature is higher at lower frequencies changing from $10^9$K at 80MHz and 169MHz to $10^{10}$K at 43MHz [8, 9]. Observations with the Nancay radio heliograph at 169MHz showed a radiation pulsation in moving Type IV bursts with the period of around 2s. This kind of pulsation was also found in Zurich at frequencies 140-250MHz during 5 min [9]. Sometimes Type IV bursts have fine structure in the form of bursts with intermediate frequency (so called fiber bursts) in emission and absorption and zebra-structure [10-12]

In the early days a synchrotron emission mechanism was used to explain the Type-IV bursts [1]. Later gyrosynchrotron mechanism was assumed [13, 14] and finally a plasma mechanism was proposed [15]. The full variety of observational data, including high emission brightness temperature, high polarization degree and its temporal variation, still cannot be completely understood within the frame of only one emission mechanism [16].

The existence of fine structure in Type IV emission is explained by different mechanisms. So, fiber bursts are supposed to appear due to coalescence of whistlers and Langmuir waves [5]. Concerning zebra-structure, the generation of Langmuir waves in double plasma resonance with their subsequent transformation into transverse waves [11, 17], or the Langmuir waves scattering on structured whistler field [11, 18] are analyzed.

In this chapter the results of observations of Type IV bursts obtained at the radio telescopes UTR-2 and URAN-2 in the frequency band 10-30MHz are considered. The observations with unprecedented sensitivity, high temporal resolution in wide instantaneous bandwidth allowed to detect Type IV bursts for the first time at such low frequencies as well as to find new properties of these bursts. These new results may appear to be the key point in understanding this kind of event.

**Observations.**

Type IV bursts discussed in the chapter were registered at the radio telescopes UTR-2 (Fig.1) (in the years 2003-2006) and URAN-2 (Fig.2) (in 2003, 2006). Three sections of the radio telescope UTR-2 with total collecting area $30000m^2$ were used providing a beaming of $1° \times 13°$. All observations were carried out with the 60-channel filter bank spectrometer. Each of 60 channels has bandwidth of 3kHz and is tuned to a fixed frequency in the range between 10 and 30MHz. The channels are spaced in average by 340kHz, but not regularly in all frequency band. In practice to minimize interference, the channels are tuned out off the interference frequency. One must take into account that on dynamic spectra the gaps between adjacent channels are filled with the values of the channels, nearest to the gap. This technique allows to clarify the spectrum and makes it more smooth along the frequency axis.

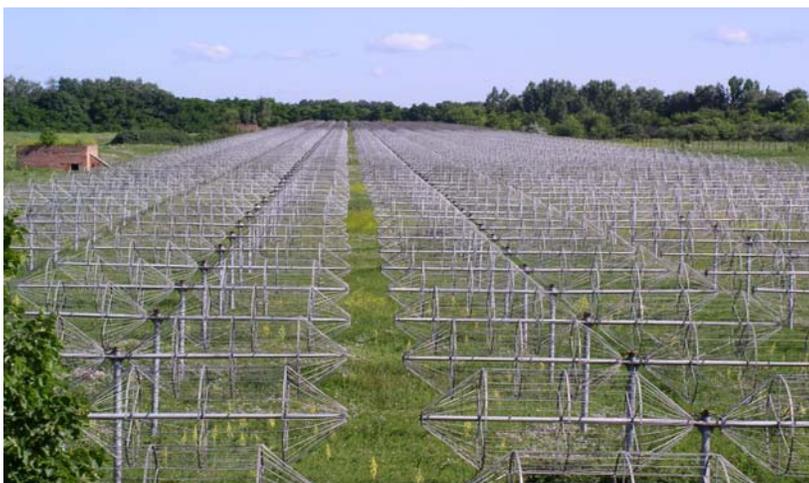

Figure 1. Radio telescope UTR-2 (Kharkov, Ukraine).

The circular polarization was measured with the radio telescope URAN-2 (Fig.2), which consists of 512 wideband (10-30MHz) dipoles [19]. The effective area of the URAN-2 antenna is $28000 m^2$, with a beaming of $3.7° \times 7°$ at a frequency 25MHz. In 2004 the polarimeter worked at two frequencies 24.5MHz and 25MHz in the frequency band of 10kHz and with a time resolution of 12.5ms.

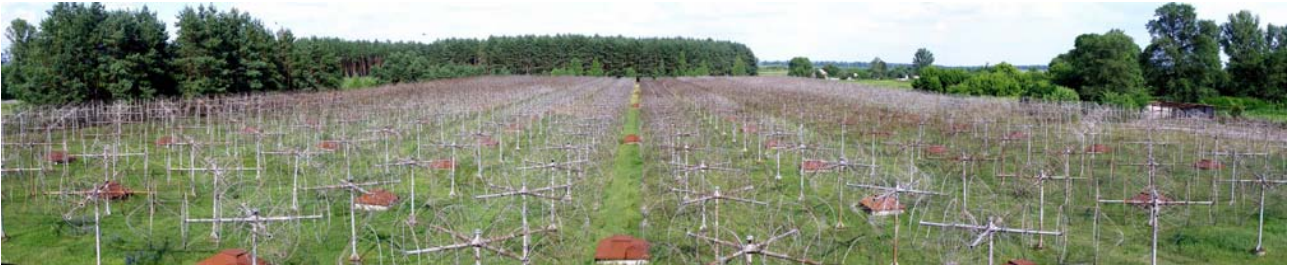

Figure 2. Radio telescope URAN-2 (Poltava, Ukraine).

During the period 2003-2006 we registered 17 events, which could be referred to Type IV bursts. In this paper we discuss only 13 of them because each considered has some important features for the understanding of Type IV burst phenomenon. For the description of decameter Type IV bursts we consider also attendant data by WIND, SOHO and Nancay Decameter Array (NDA) if they exist.

*On June 13, 2003*

This Type IV burst began over the frequency range 10-30MHz nearly at 07:00 UT (Fig.3). The time profile of the burst is characterized by fast rise (approximately 1 hour) and slower decay (about 3h 30min practically till 11:30 UT) (Fig.4). The front edge of the burst drifts from high to low frequencies with the rate of 10kHz/s. The rear part of the burst, apparently, drifts with the same drift rate (observations with NDA at higher frequencies are also in favor of this (Fig.5)). On a dynamic spectrum below 20MHz one can see a shading area, the front edge of which has positive frequency drift rate and the rear edge, on the contrary, has a negative one. On the time profile of the burst in addition to the main maximum (with flux reaching 1000s.f.u.) there can be seen also the second and probably the third maximum (with fluxes 20s.f.u. and 10s.f.u., respectively). The frequency profile at maximum phase of the burst (at 08:20 UT) (Fig.6) shows fast fall (from 600s.f.u. at 30 MHz to 2s.f.u. at 10 MHz) from high frequencies to low frequencies.

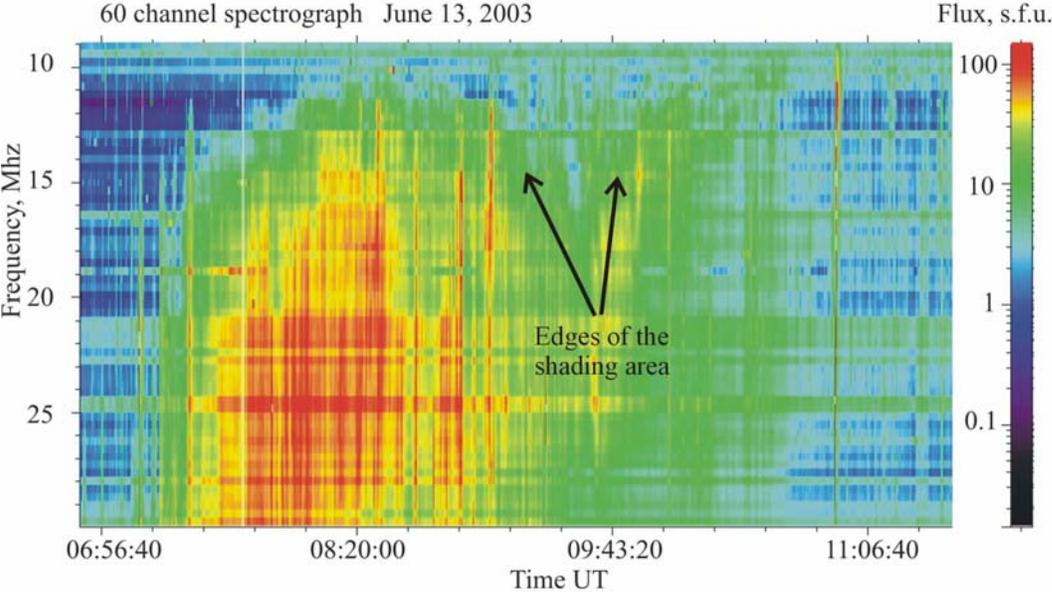

Figure 3. Type IV burst, which moves with drift rate 10kHz/s from high to low frequencies.

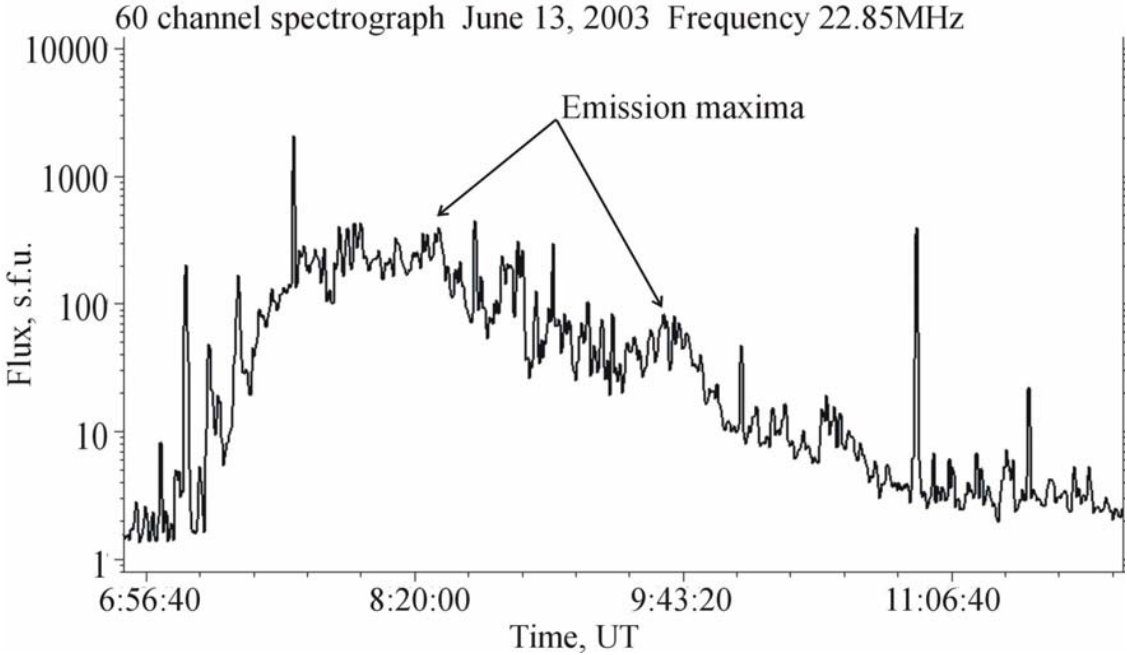

Figure 4 Time profile of Type IV burst. There is the second maximum[h1] at 09:46:20 UT.

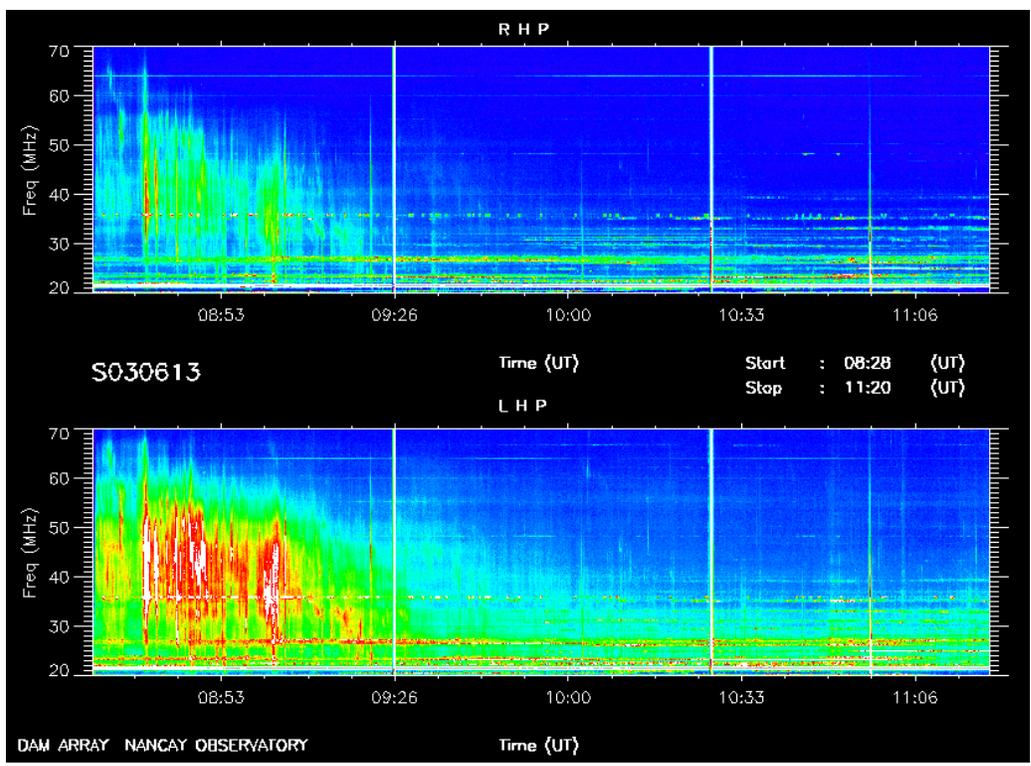

Figure 5. Right and left hand polarization of Type IV burst observed by NDA

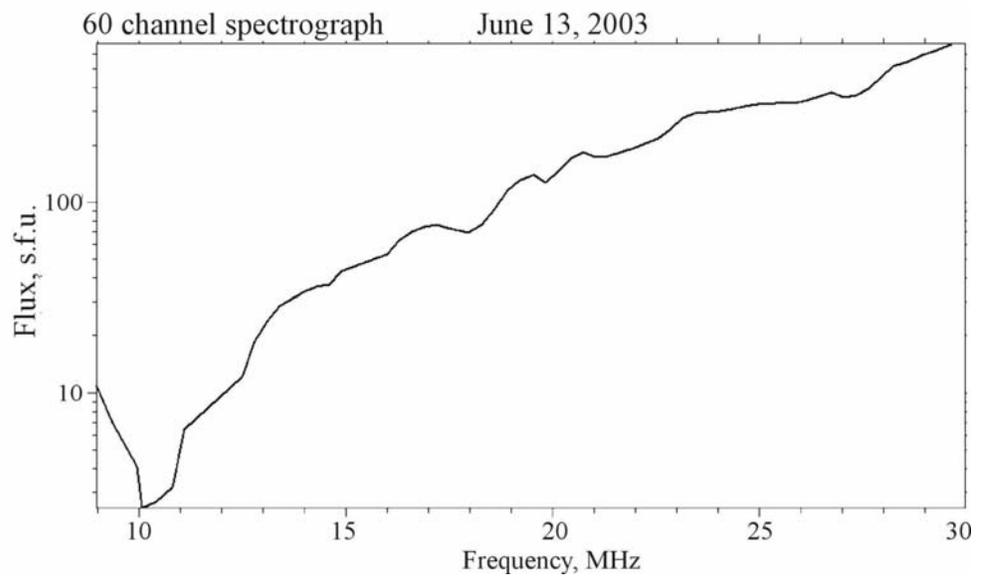

Figure 6. Frequency profile of Type IV burst maximum phase (at 08:20 UT)

As in all observed Type IV bursts, the considered burst had fine structure in the form of bursts, which in the overwhelming majority drifted from high to low frequencies with rates about 1-2MHz/s, slightly lower than the drifts of common Type Ш bursts for this frequency band. The duration of these sub-bursts was 10-20 s, which is more, than for common Type III bursts. It is necessary to note also, that at the beginning of the Type IV burst the sub-bursts are less observable than at the end of it. The maximum fluxes of sub-bursts with respect to the continuum flux are comprising in the majority only 10-20 % . Some fast drifting sub-bursts were also registered. Their drift rates were essentially greater, with regard to the rest of sub-bursts, and in rare cases the drift rates were positive. This Type IV burst is one of the three bursts discussed for which there are the simultaneous observations of polarization obtained by the radio telescope URAN-2 (Poltava). Fig.7 shows the evolution in time of the polarization degree, since 07:28 UT till 10:57 UT. It can be seen that the polarization degree gradually grows, reaching the maximum value of 55% at 09:40 UT, and then starts to fall practically down to 10%. Judging from the heliographic observations by Nancay (Fig.8), this burst is connected to the active regions situated on the western part of the solar disk.

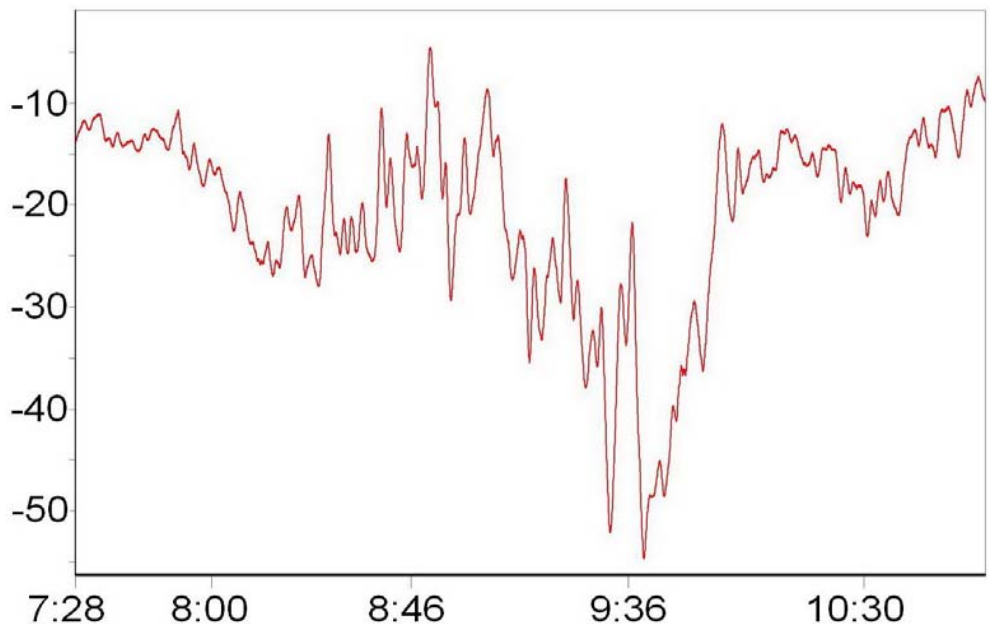

Figure 7. Polarization of Type IV burst measured by URAN-2.

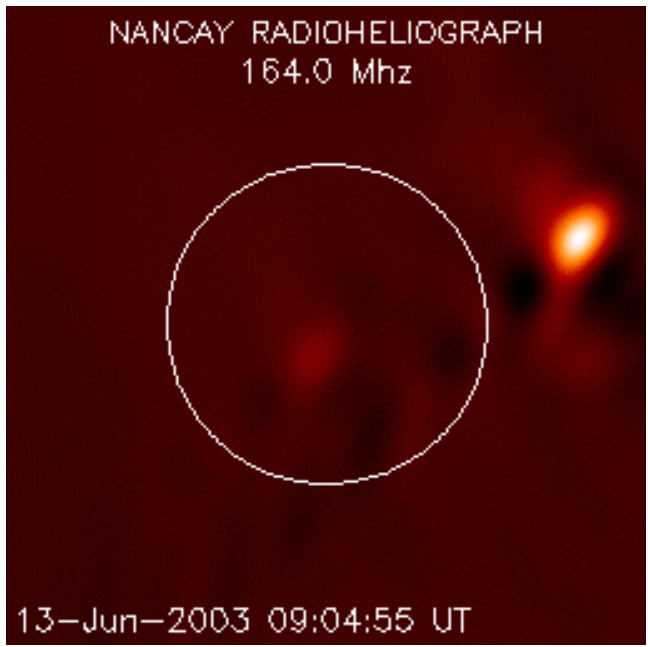

Figure 8. Position of source of 164 MHz radio emission near the maximum phase of Type IV burst.

*On July 22, 2003*

This Type IV burst is characterized by rather slow intensity rise. Starting at approximately 10:00 UT it reaches a maximum (100s.f.u.) only at 11.00 UT (Fig. 9). This burst also has at least two feebly marked increases of the emission intensity (Fig. 10). The internal structure of this burst has an appearance of well separated bursts similar to Type III bursts. Though the durations of these sub-bursts are 5-10s which is practically the same as for classical Type III bursts, their drift rates are in the majority less than 1MHz/s which is slower than usual Type III bursts exhibit. As it can be seen from the Fig.11 the Type IV burst consists also of Type IIIb bursts. In general these Type IIIb bursts have the same durations and drift rates as they usually have during Type IIIb storms, but in some cases the drift rates are less than 1MHz/s. A remarkable feature of the instant spectrum of this Type IV burst (taken at maximum flux) is the flux rise towards the lower frequencies (Fig. 12). Such flux versus frequency dependence is opposite to that found for the majority of the observed Type IV bursts. The measured polarization of the burst, which is shown in Fig. 13, is rather small. This Type IV burst fills an intermediate place between Type III storm and the Type IV burst itself.

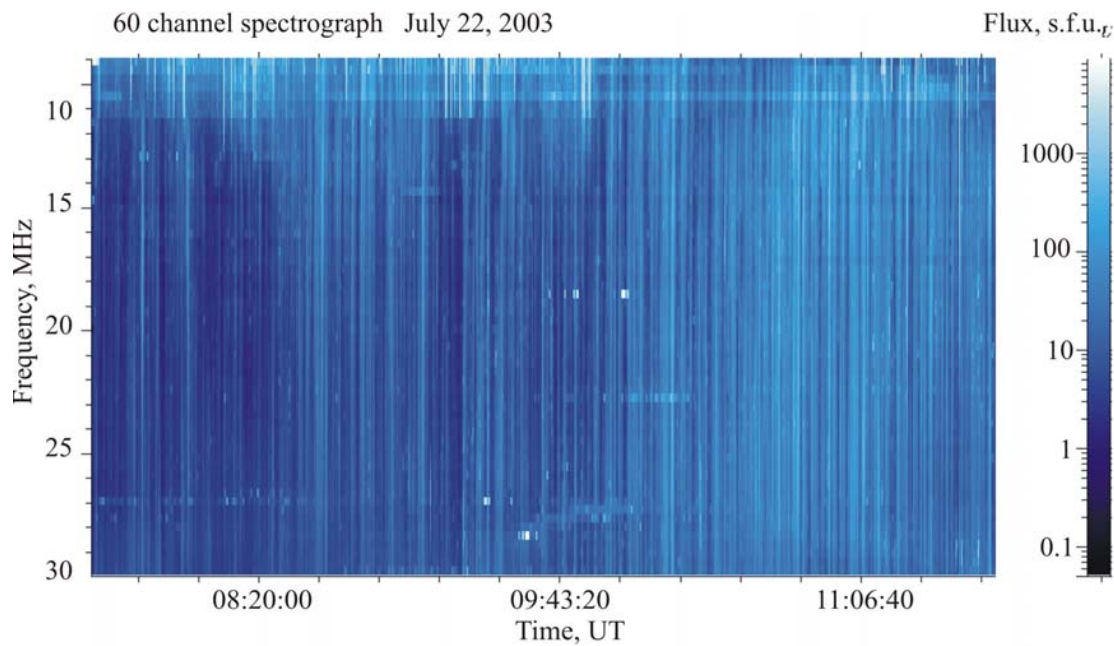

Figure 9. Dynamic spectrum of solar radio emission on July 22, 2003, observed with UTR-2.

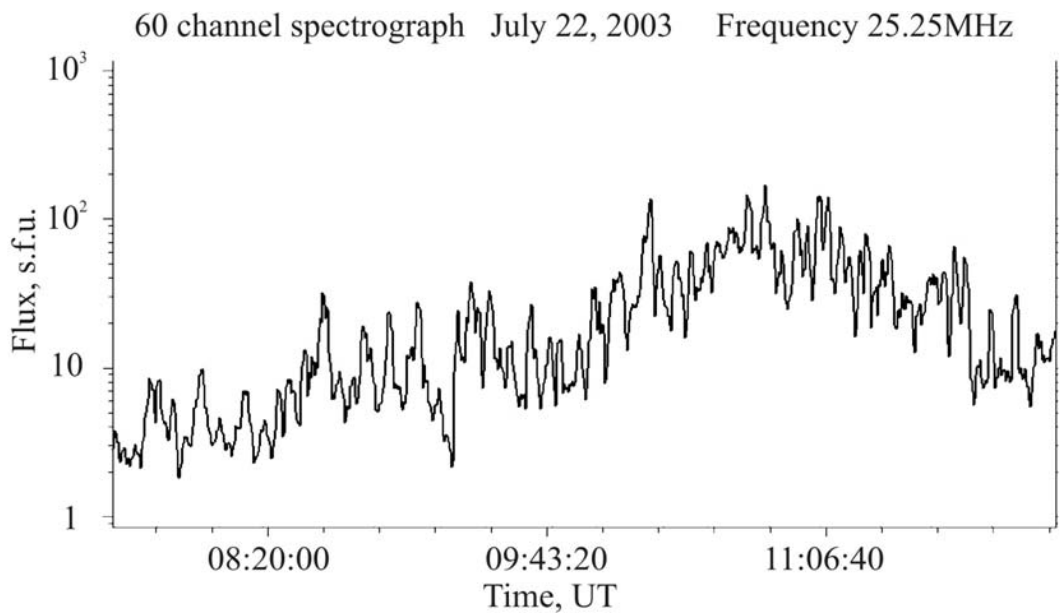

Figure 10. Time profile of Type IV burst with slow increase and maximum at 11:06 UT.

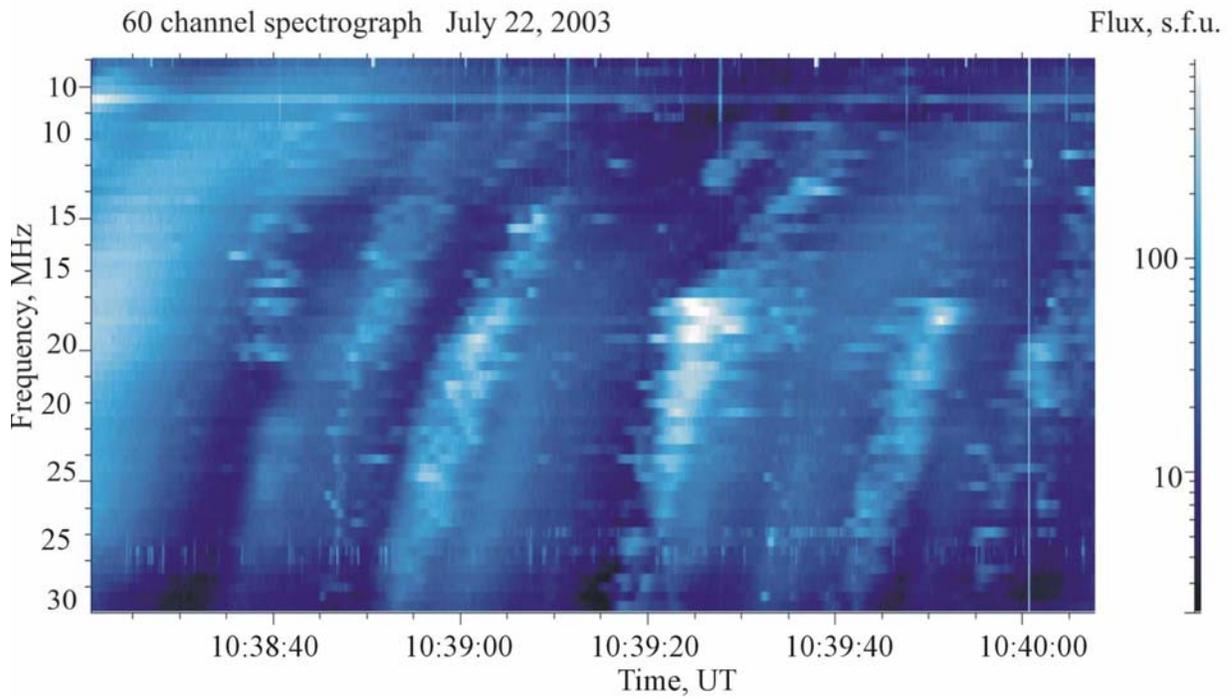

Figure 11. Fragment of Type IV burst with fine structure in the form of bursts similar to usual Type III, Type IIIb bursts, but unusual low drift rates (<1MHz/s).

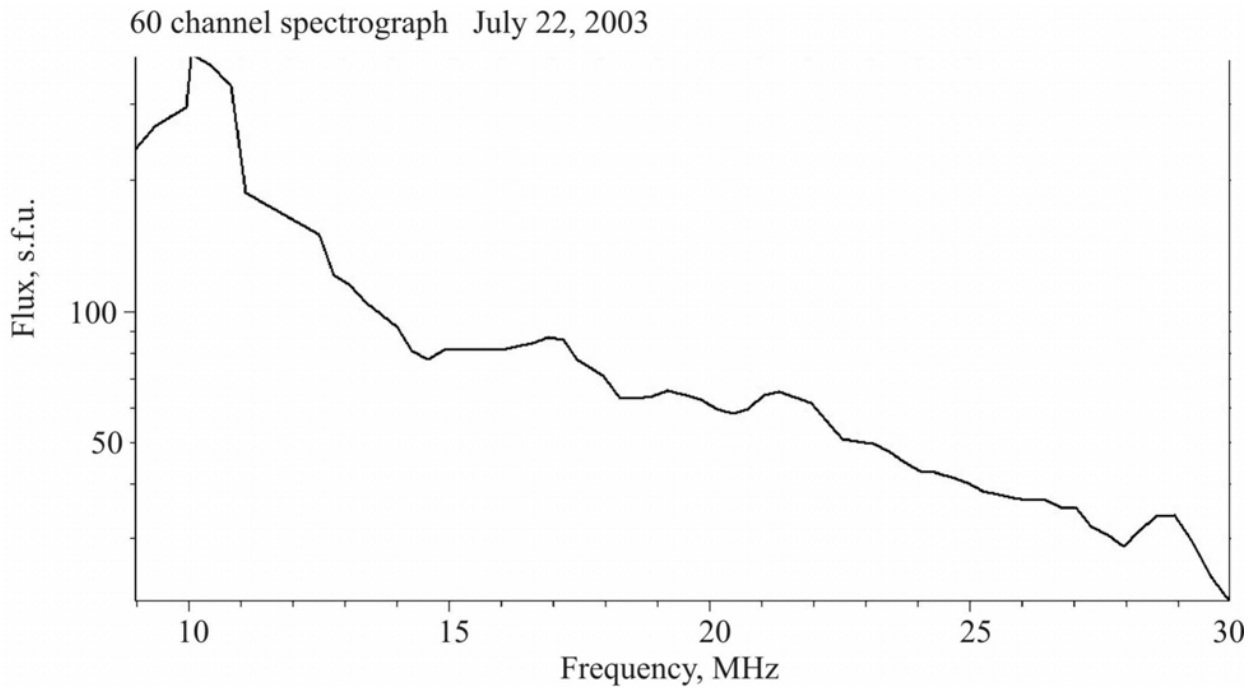

Figure 12. Frequency profile of Type IV burst at maximum phase (11:06 UT)

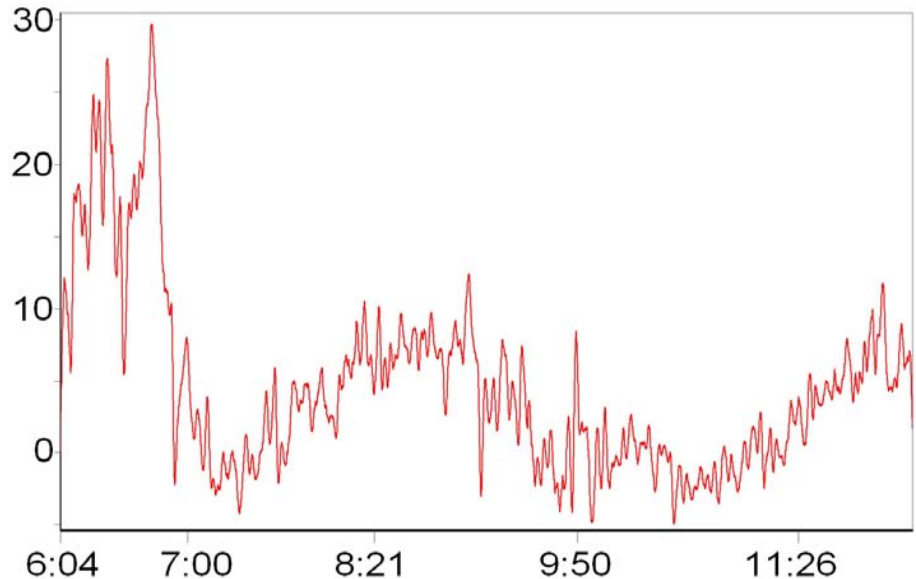

Figure 13. Nonmonotonic change of radio emission polarization at 25MHz during Type IV burst

*On August 3, 2003*

This Type IV burst is not intensive - its maximal flux is about 20s.f.u. (Fig.14). It has a symmetric time profile with full duration of about 2 hours and a maximum intensity falling at 11:00 UT (Fig.15). The burst is observed on a background of powerful Type III bursts with fluxes reaching up to 10000s.f.u. The dependence of flux via frequency for observed Type IV burst with 100s averaging is shown in Fig.16. It is seen that the flux grows with falling frequency, as it was for the event on July 22, 2003, but conversely to the flux-frequency dependence for the 13 June 2003 event. It is interesting, that averaging the flux with 100s averaging time for all observation time on this day gives also the second maximum, which falls at 8:20 UT (Fig.16). Here the main contribution to the radiation intensity is provided by a group of intense Type III bursts. These two periods of the increased radio emission, apparently, reflect the activity of active regions. And if in case of the activity, which occurs between 08:00 UT and 09:15 UT, the increased level of radiation was provoked mainly by accelerated electron beams emitted from the active region, in the second case there was a smooth plasma heating producing the increased continuum radiation of the Type IV burst. The fine structure of the Type IV burst in the majority has an appearance of Type III bursts with their inherent durations and frequency drift rates observed on a continuous background.

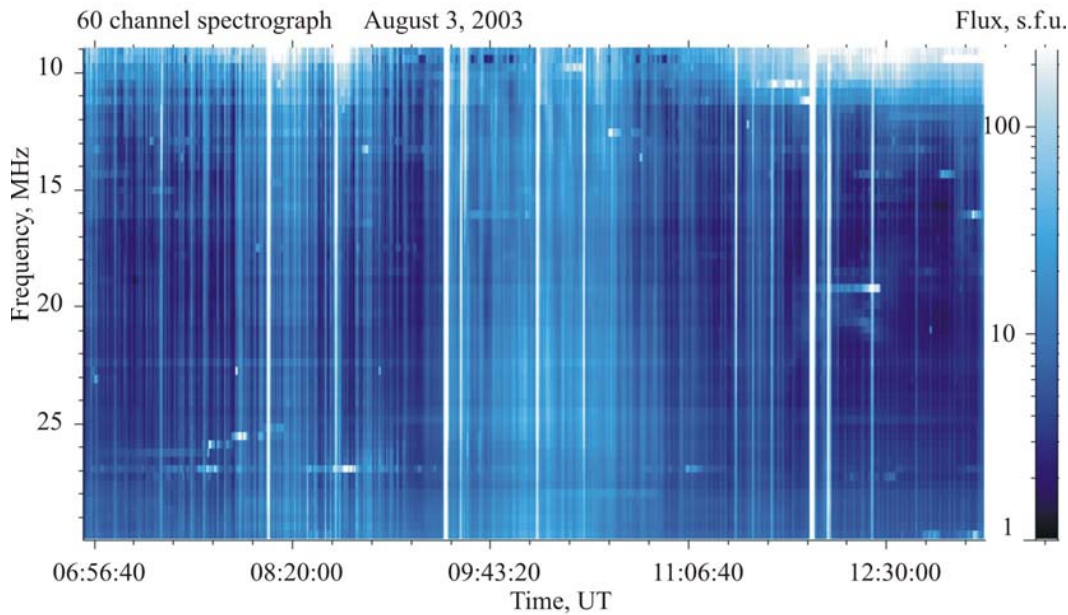

Figure 14. Weak stationary Type IV burst (from 09:30 UT to 11:20 UT).

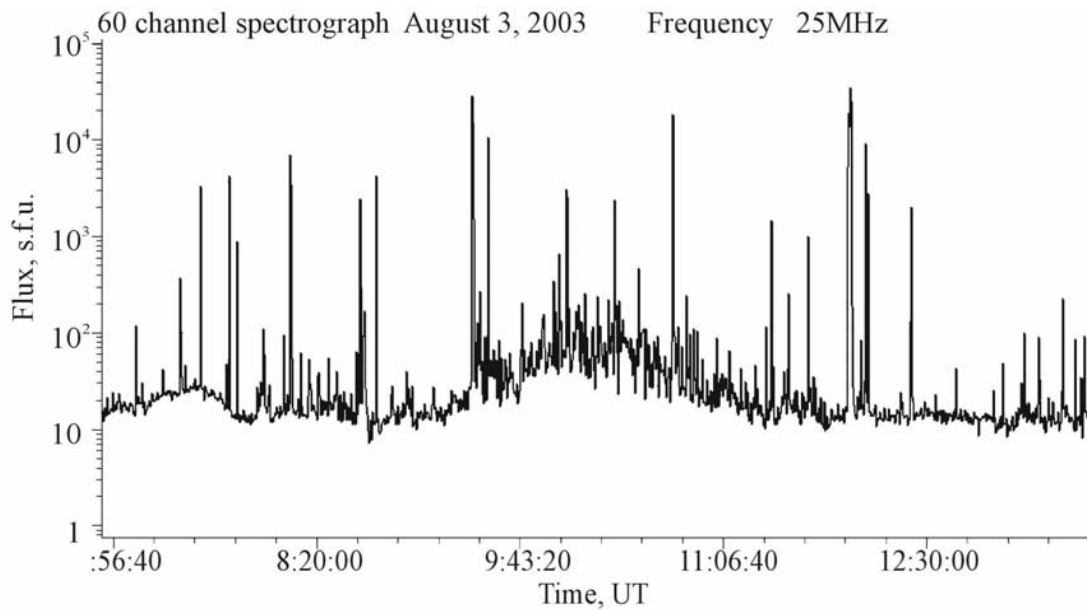

Figure 15. Time profile of solar radio emission on August 3, 2003 at 25MHz without time averaging (time resolution in these observations was 50ms)

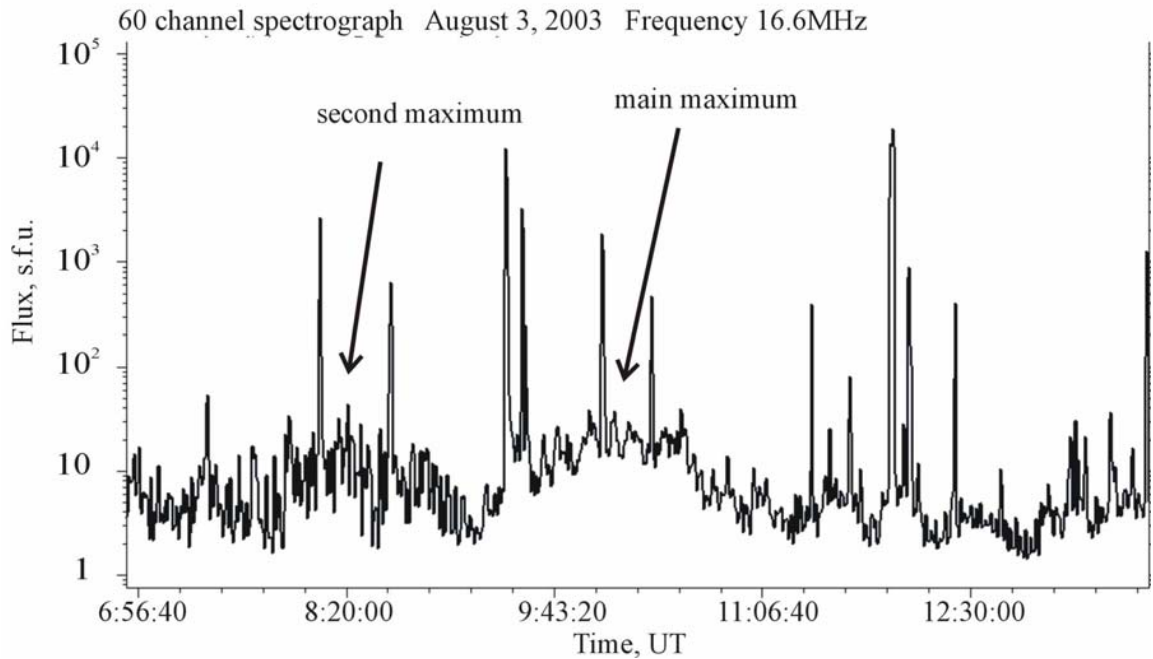

Figure 16. Time profile of solar radio emission on August 3, 2003 at 16MHz with 100s averaging.

*On August 19, 2003*

At 07:55 UT the group of very powerful Type III bursts with fluxes of more than 10000s.f.u. has been seen (Fig.17). Right after this a group of fiber bursts with drift rates 30kHz/s was observed (Fig.18). Their durations were up to 3s, and fluxes reach 300s.f.u. Later on the Type II burst starts drifting from high to low frequencies. This burst appeared to consist of two and probably even three harmonics with starting times 08:09, 08:16 and 08:19 UT. The first and the second harmonic had drift rates of about 40-50kHz/s and were splitted into two lanes each. Each of the Type II burst lanes consists of sub-bursts with the drift rates 1 - 4MHz/s and durations equal to 3-5s. The emission of an increased level, which is observed right after the group of intense Type III bursts steeply falls off reaching a value of 0.5s.f.u. by 08:30 UT. At low frequencies (near 10MHz) there can be seen a slight rise of emission between 09:15 UT and 10:15 UT. At 10:05 UT the lanes of Type II burst start at high frequencies (Fig.19). But in this case their drift rates equal 20kHz/s. Right after that, at 10:23 UT at 25MHz, another type II burst is registered. It has the same drift rate and is a little bit weaker. It is superimposed by two groups of Type III bursts with the peak fluxes of 500-600s.f.u. Just after them the Type IV burst begins.

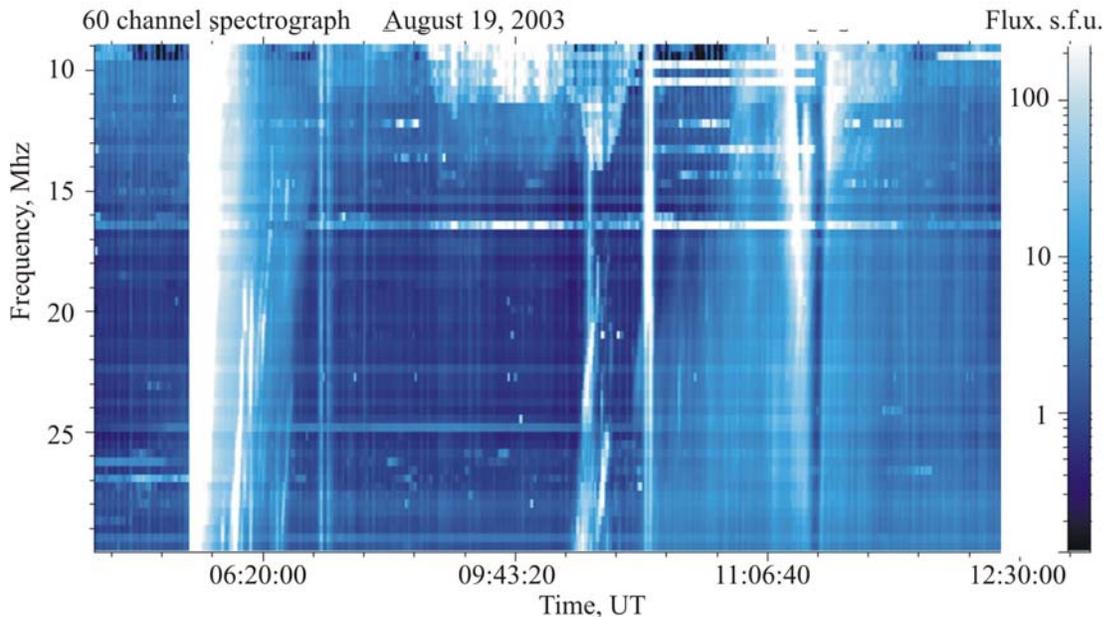

Figure 17. Type IV burst (from 10:25 UT to 12:30 UT) with attendant phenomenon group of Type III bursts, Type II bursts.

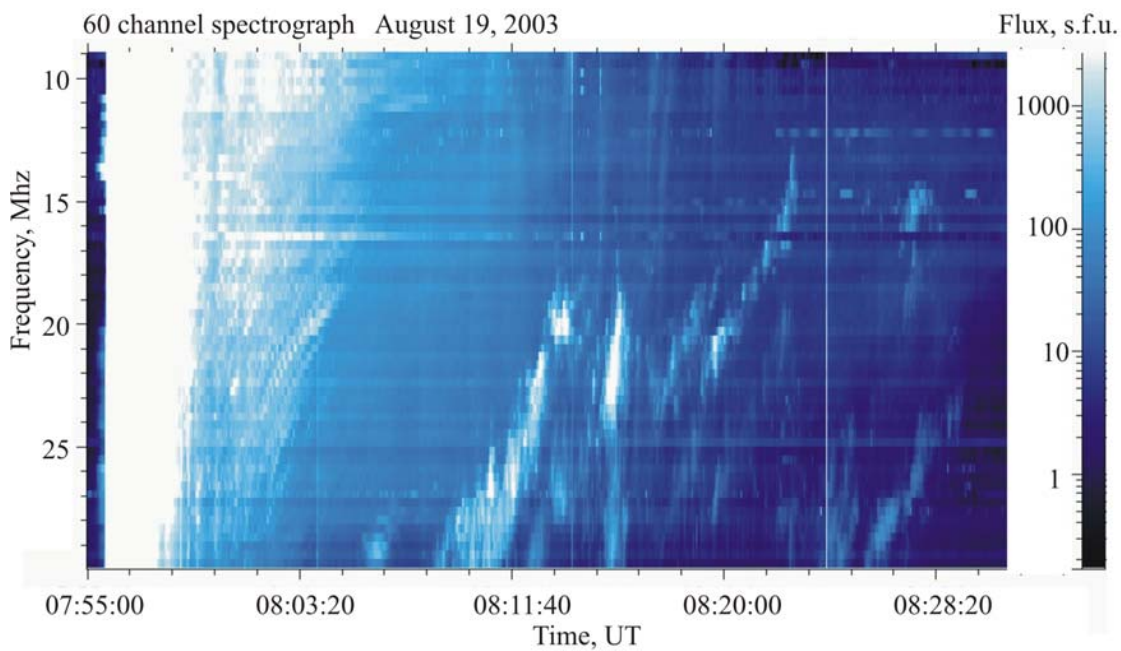

Figure 18. Group of Type III bursts at 07:55 UT, fiber bursts directly after it, fundamental and harmonic Type II burst.

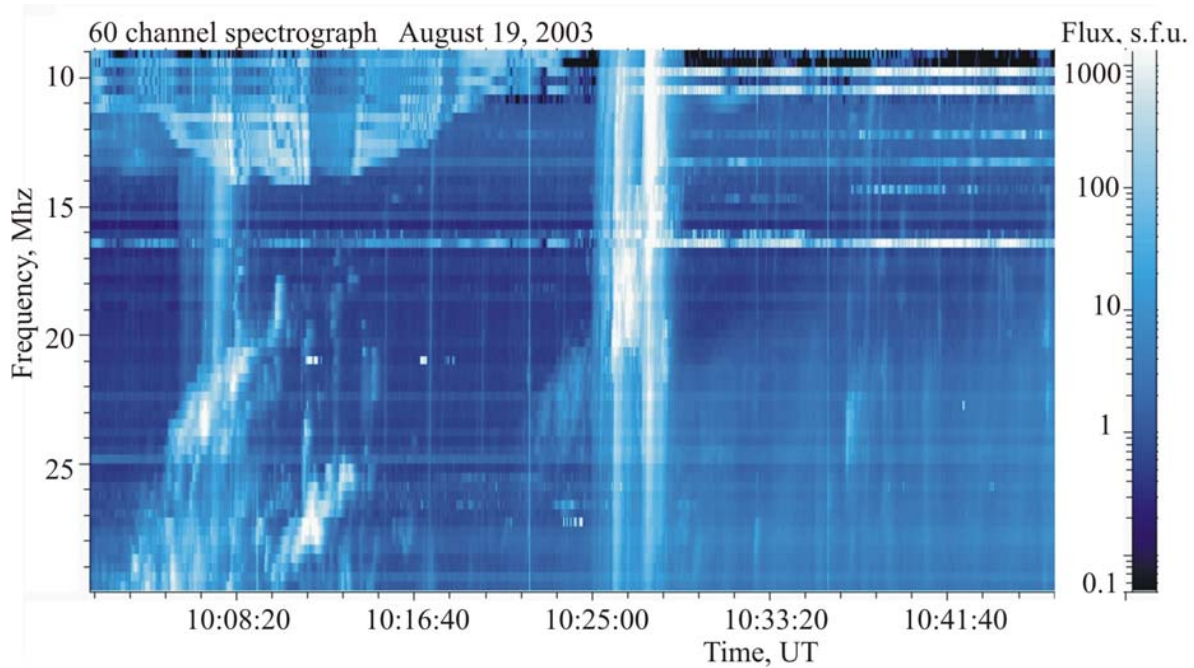

Figure 19. Type II burst (with lanes), group of Type III bursts and beginning of Type IV burst.

The forward edge of the burst drifts slowly down to low frequencies with 10 kHz/s drift rate. The flux reaches its maximal value 30s.f.u. at 11:10 UT. Thus, rise time of this Type IV burst amounts roughly to 40min. The tail of the burst is rather long - by the time 12:45 UT (the end time of observations on that day) the emission level is still higher than 1s.f.u. The second, less pronounced emission maximum can also be found at 12:05 UT. The Type IV burst consists of very weak sub-bursts, which are hardly distinguished on a continuum background. The drift rates of these sub-bursts equal mainly 1-2MHz/s, and durations are 10-20s. At 11:17 UT the absorbing area, which widens towards low frequencies is observed below 15 MHz (Fig.20). Well visible in all frequency band from 10 to 30MHz burst in absorption starts at 11:23:20 UT (Fig.20). It drifts from high to low frequencies with drift rate of 120kHz/s. The total duration of the burst in absorption equals 5min 20s. The minimal flux at 25MHz amounts to 0.6s.f.u., which exceeds the solar emission level before Type IV burst (Fig.21). The bursts in absorption are observed on a background of fiber bursts, which drift from high to low frequencies with velocities 30kHz/s. Their durations equal 2-3s (this burst is discussed in detail in [20.21].

According to the SOHO data (Fig.22) the powerful CME has been registered at 10:24 UT, which moves with 468km/s. By the time of Type IV burst maximum (11:00 UT) the CME front was at a distance of approximately $4R_s$ from the solar surface.

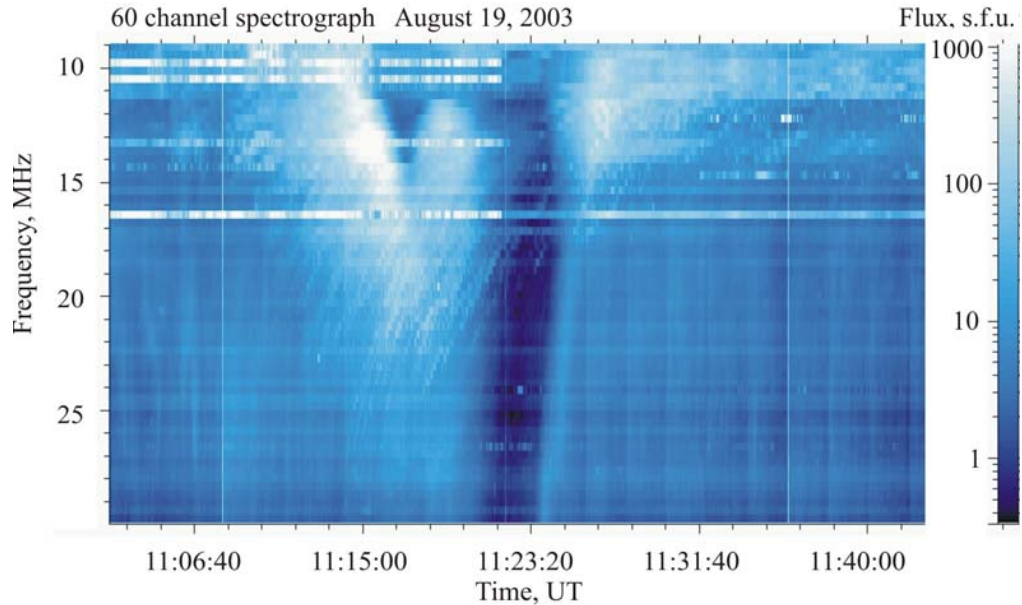

Figure 20. Absorbing area (at low frequencies at 11:17 UT) and burst in absorption against fiber bursts.

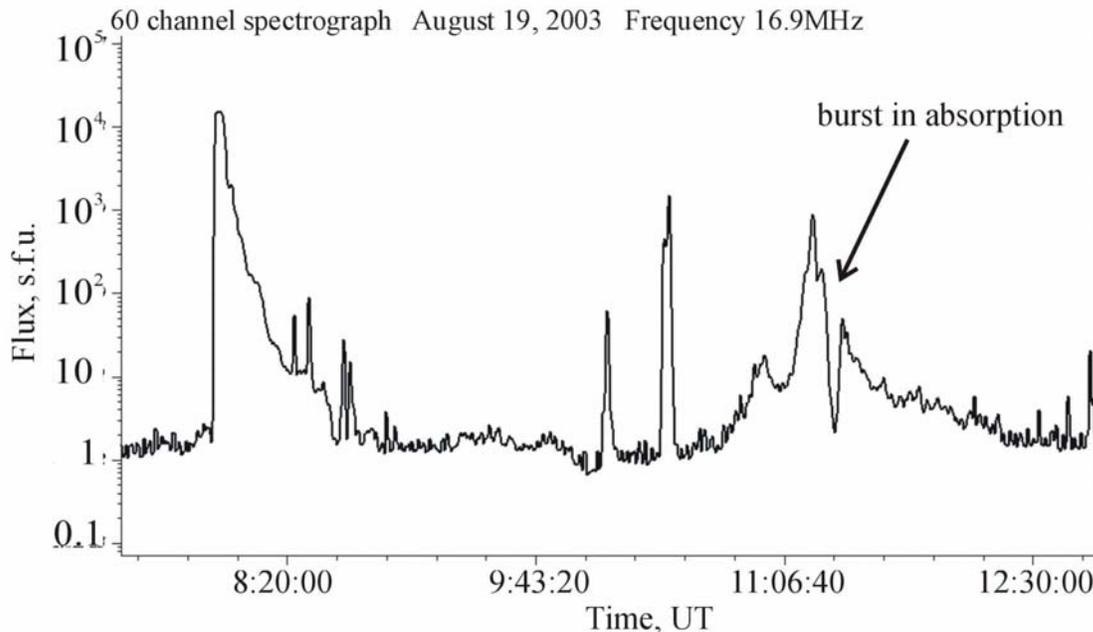

Figure 21. Time profile of solar radio emission on August 19, 2003. There is clearly seen the burst in absorption [h2]at 11:20 UT.

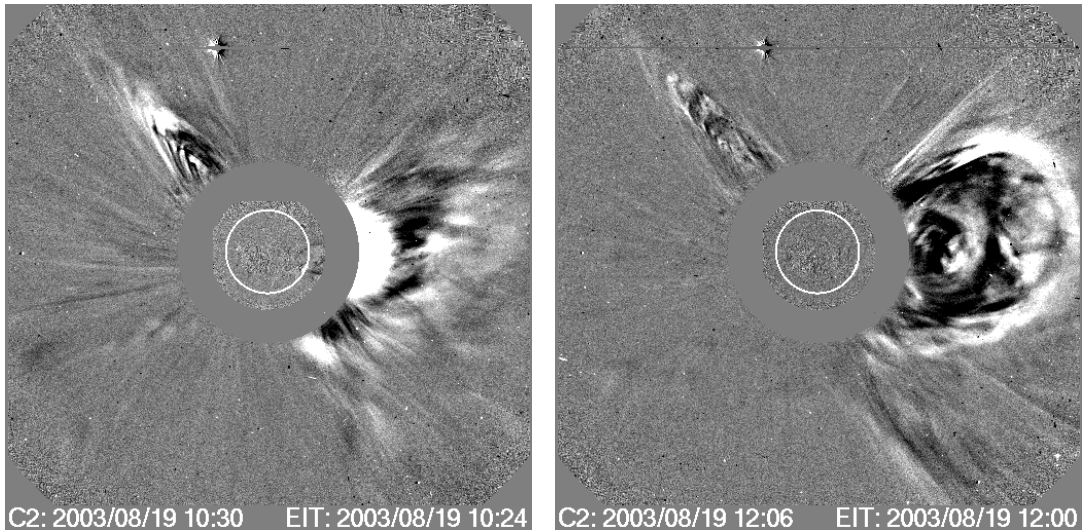

Figure 22. Movement of west CME associated with August 19, 2003 Type IV burst according to SOHO data.

*On June 4, 2004*

This event (Fig. 23), as well as the event on August 19, 2003, starts with a group of intense Type III bursts observed at 07:33 UT. The maximum fluxes reach 3000s.f.u. The bursts durations are about 10s, and their drift rates are 2-5MHz/s. At the end of this group one can see some fiber bursts with drift rates of 90kHz/s (Fig.24). Their durations are the same as for the event on  August 19, 2003, namely 3-5s. Then the burst in absorption is observed. Its duration strongly depends on frequency varying from 2min 30s at 30MHz to 10min 50s at 10MHz. The front of the burst in absorption has practically infinite drift rate, and the tail is limited by the first harmonic of Type II burst. The latter drifts with the rate of about 40kHz/s.

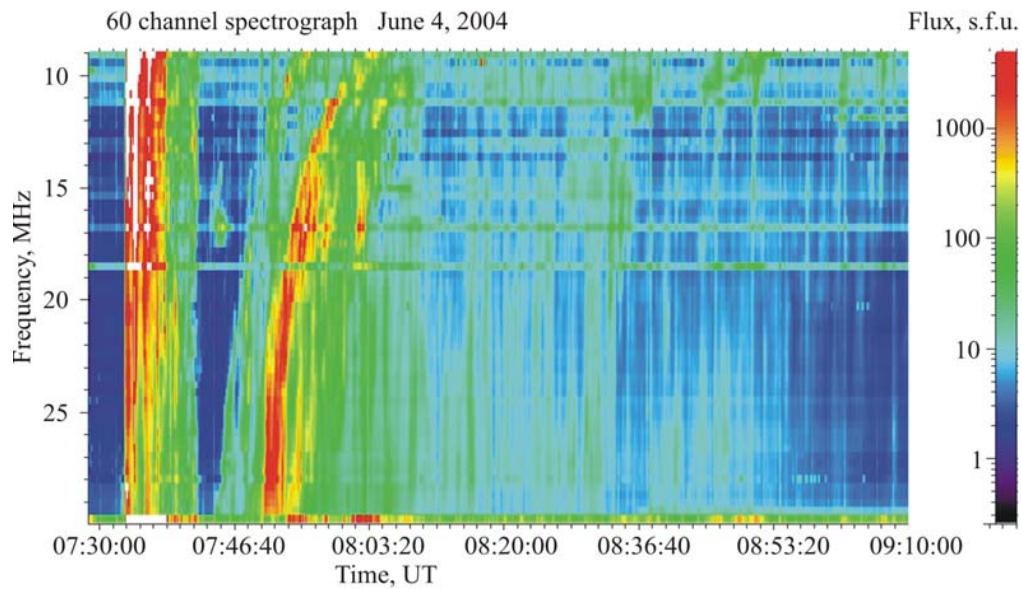

Figure 23. Type IV burst as succession of Type III bursts group, burst in absorption, and Type II burst.

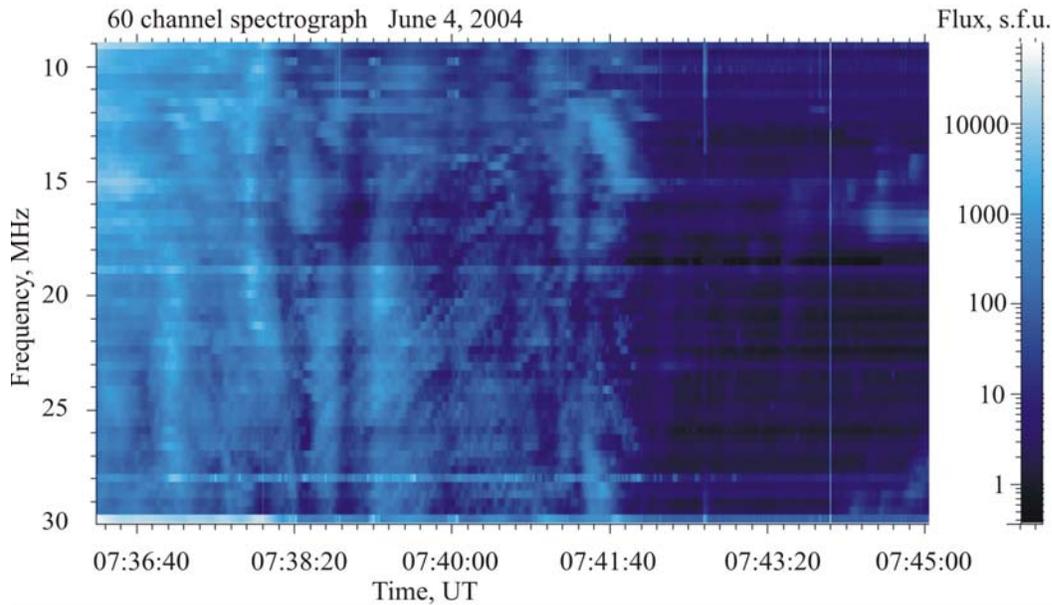

Figure 24. Fiber burst with drift rates about 90MHz/s directly after a group of intensive Type III bursts.

A remarkable feature of this Type II burst is the presence of the first, second and third harmonics (Fig.25). The second harmonic is the most intense, and the third one is the weakest. All harmonics are splitted into lanes and have fine structures in the form of sub-bursts with  for decameter wavelengths standard durations (5-10s) and drift rates (1-5MHz/s). These three harmonics are also well distinguished on the dynamic spectrum obtained by the WIND spacecraft (Fig. 26). The Type II burst is followed by the Type IV burst. However we can see only a falling branch of this burst (Fig. 27). The burst lasts up to the end of observations on that day - till 9:30 UT. On the falling edge of the Type IV burst there are, at least, two increases of the emission intensity: at 08:30 UT and at 08:48 UT. The Type IV burst consists of fiber bursts (Fig. 28) with drift rates less than

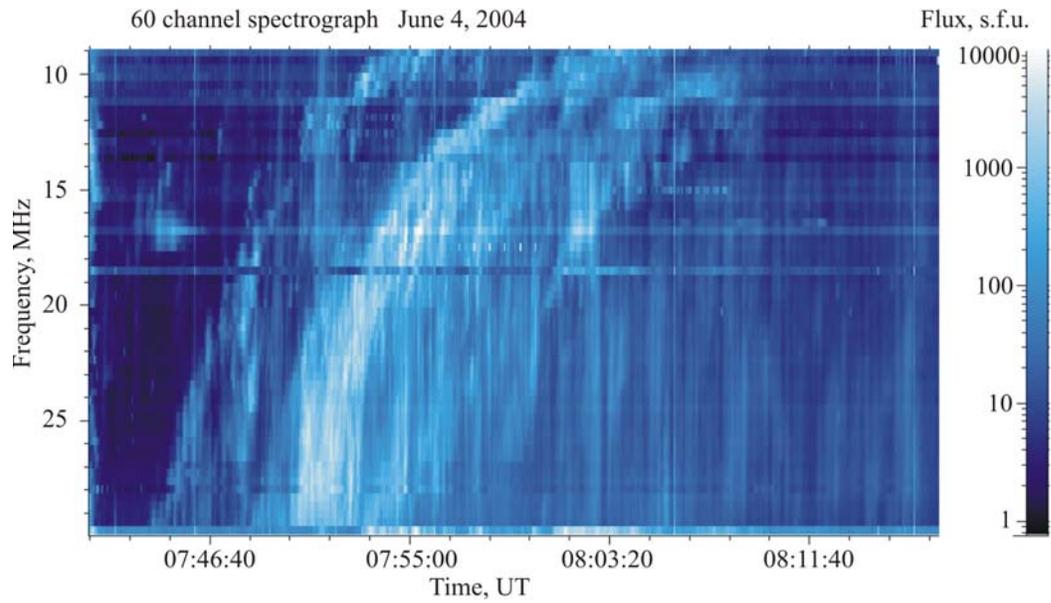

Figure 25. Type IV burst followed Type II burst with three harmonics.

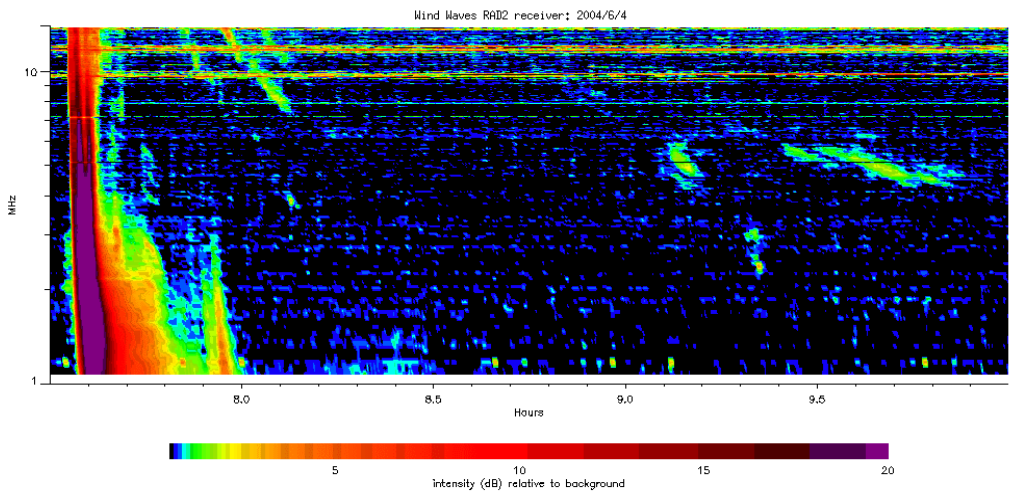

Figure 26. Type II bursts with three harmonics at frequencies 4-11MHz (WIND data).

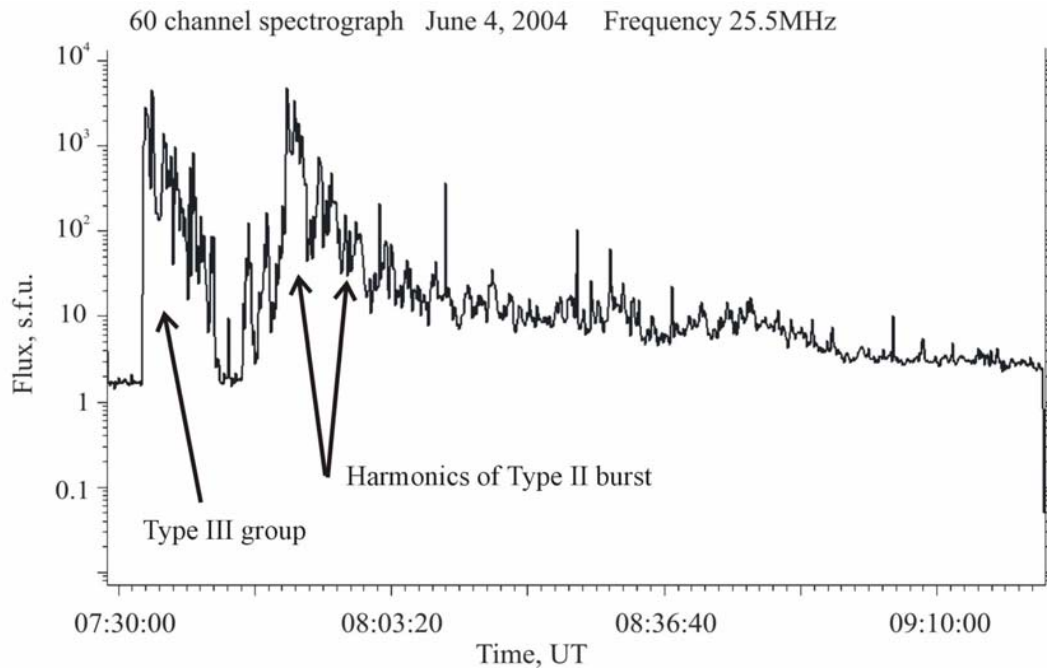

Figure 27. Time profile of June 4, 2004 event.

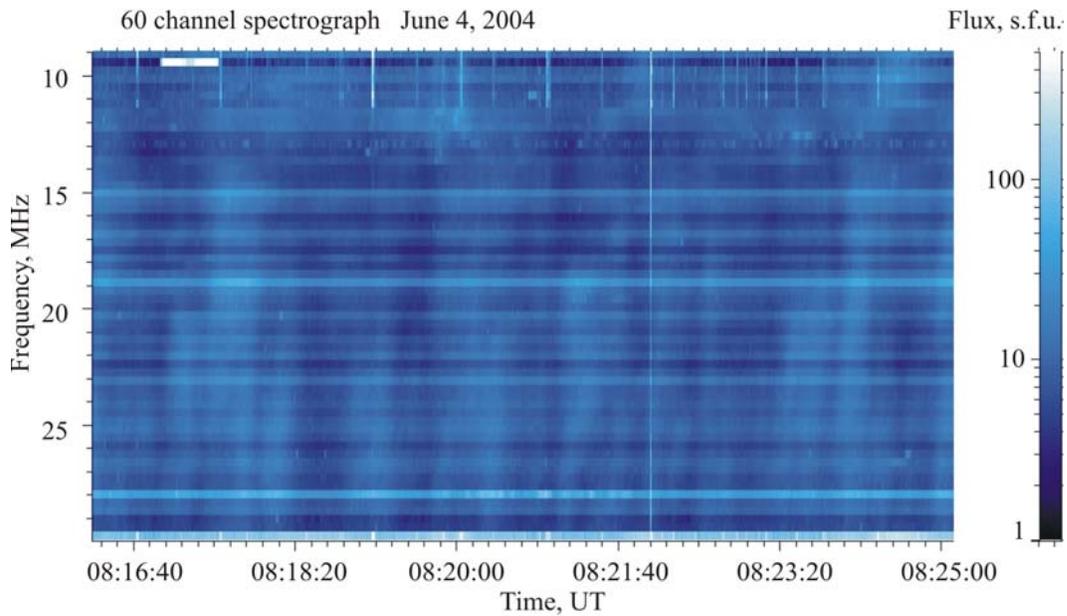

Figure 28. Fiber bursts as a fine structure of Type IV burst.

1MHz/s, and sometimes more - up to 2-3MHz/s and durations of 10-20s. Practically all fiber bursts have negative drift rates although sometimes there are also bursts with positive drift rates. At the time of occurrence of the Type II burst at the decameter band the SOHO coronograph has detected a powerful CME (Fig. 29), which moved very fast in the solar corona (according to SOHO data the CME speed was 1306km/s).

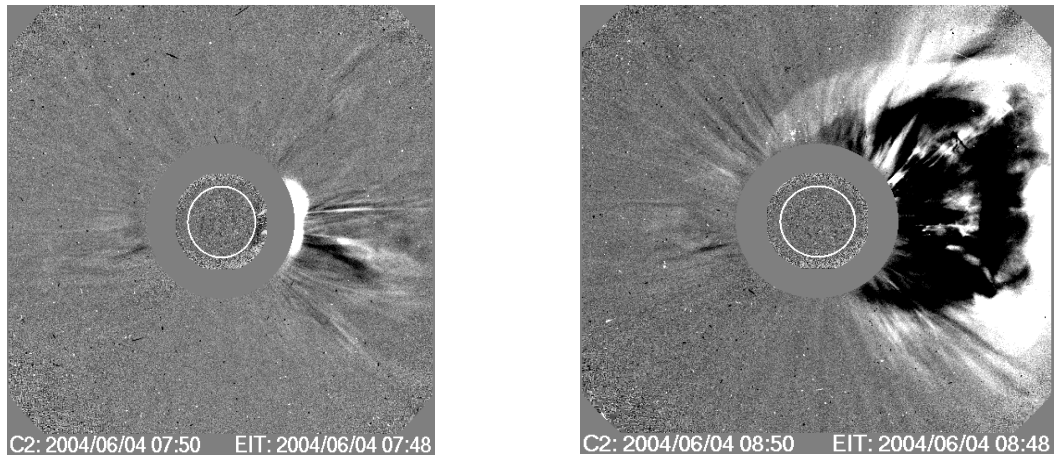

Figure 29. CME during  June 4, 2004 event

*On July 13, 2004*

At 08:46:50 UT   (Fig.30) two intensive Type III bursts were observed. Their fluxes are about 10000s.f.u. and total duration exceeds 20s. They are followed by one more group of Type IIIb-III bursts. The Type II bursts started at 08:56:20 UT at a frequency 30MHz (Fig.31). Its drift rate was about 50kHz/s.  Later on, at 09:02:20 UT the second harmonic of this burst appeared. All harmonics have fine structure in the form of drifting sub-bursts, but the second harmonic consists of the shorter sub-bursts (2-5s), with regard to the first harmonic (10-20s).

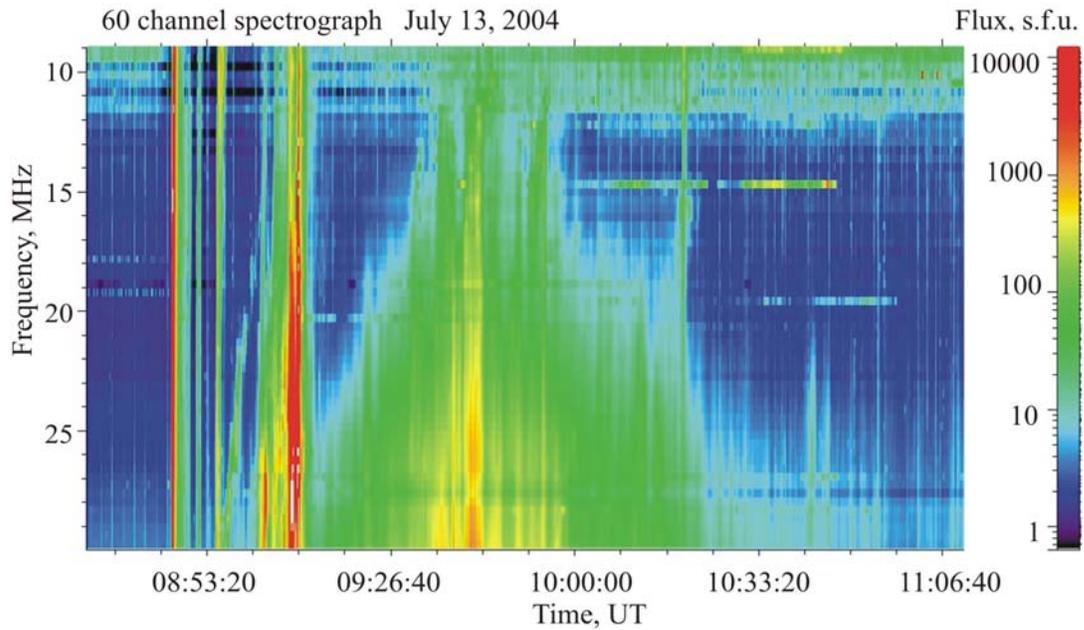

Figure 30. Low frequency part of Type IV burst with attendant Type III bursts and Type II burst.

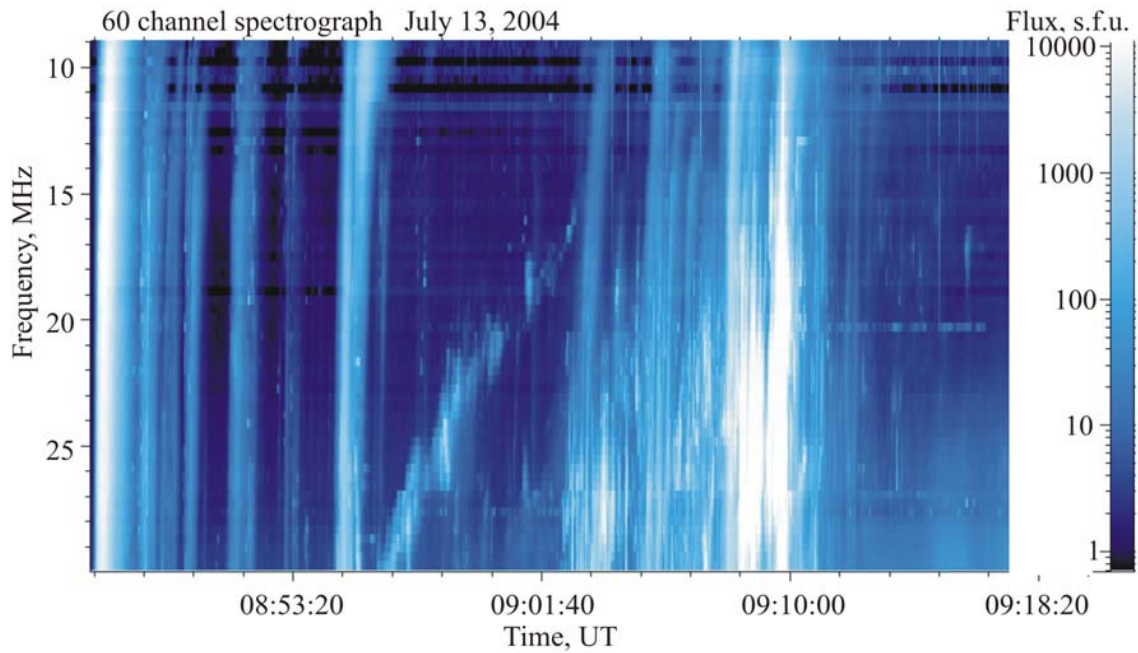

Figure 31. Group of Type III bursts and two harmonics of Type II burst intersecting with Type III bursts.

At 09:11:00 UT at high frequencies the Type IV bursts has started (Fig.32).

It also consists of sub-bursts with durations of about 5s and negative drift rates 1-3MHz/s, which are then followed by a continuum with an increased flux level of around 20s.f.u. The intensity grows rather fast and reaches 200s.f.u. at 09:45:00 UT (Fig.33).

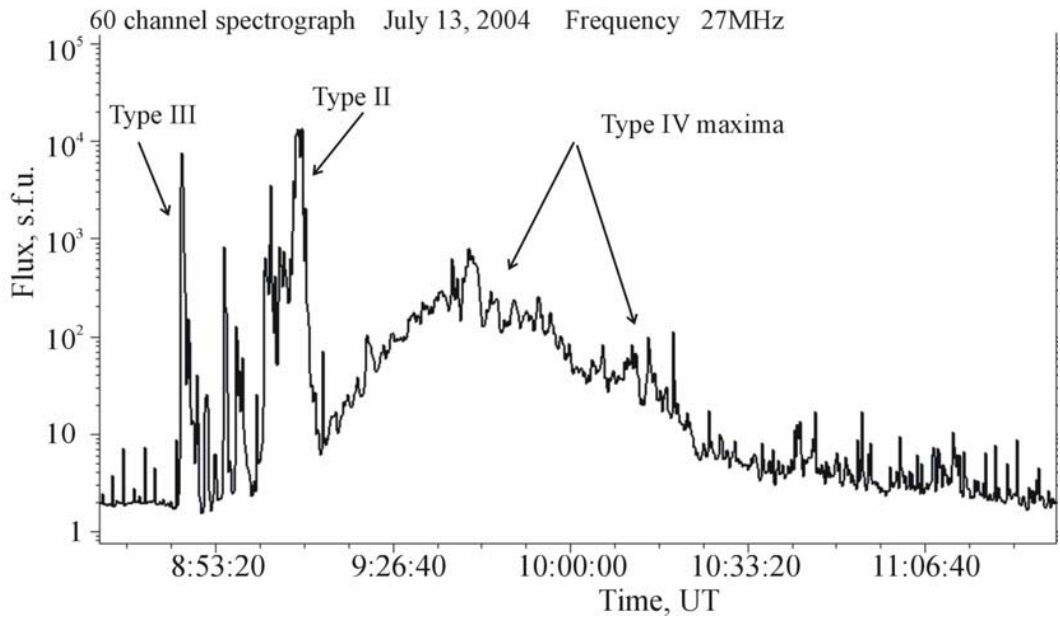

Figure 32. Time profile of Type III bursts, Type II burst and Type IV burst with two maxima.

According to our data, and data obtained from NDA (Fig.33) this Type IV burst has two components. The second component can be seen at 25MHz at 10:15:00 UT. Apparently, this burst is stationary and at frequencies 10-30MHz we observe only the lower frequency part of the burst, the main part however being at higher frequencies. The discussed burst consists of a great number of sub-bursts with various durations ranging from 4-5s up to 20s (Fig.34). The drift rates of the sub-bursts could be both negative and positive with absolute values variable from 2-3MHz/s to 10MHz/s. The sub-bursts are more distinctly seen on a falling part of the Type IV burst in spite of the fact that their intensity appreciably decreases due to increasing distance between them. As in the previous cases this burst accompanies the CME (see Fig.35) with a velocity 713km/s. The instant spectrum of the Type IV bursts at the time of the flux maximum has steep falling slope from high to low frequency changing from 300s.fu. at 30MHz down to 20s.f.u. at 10MHz.

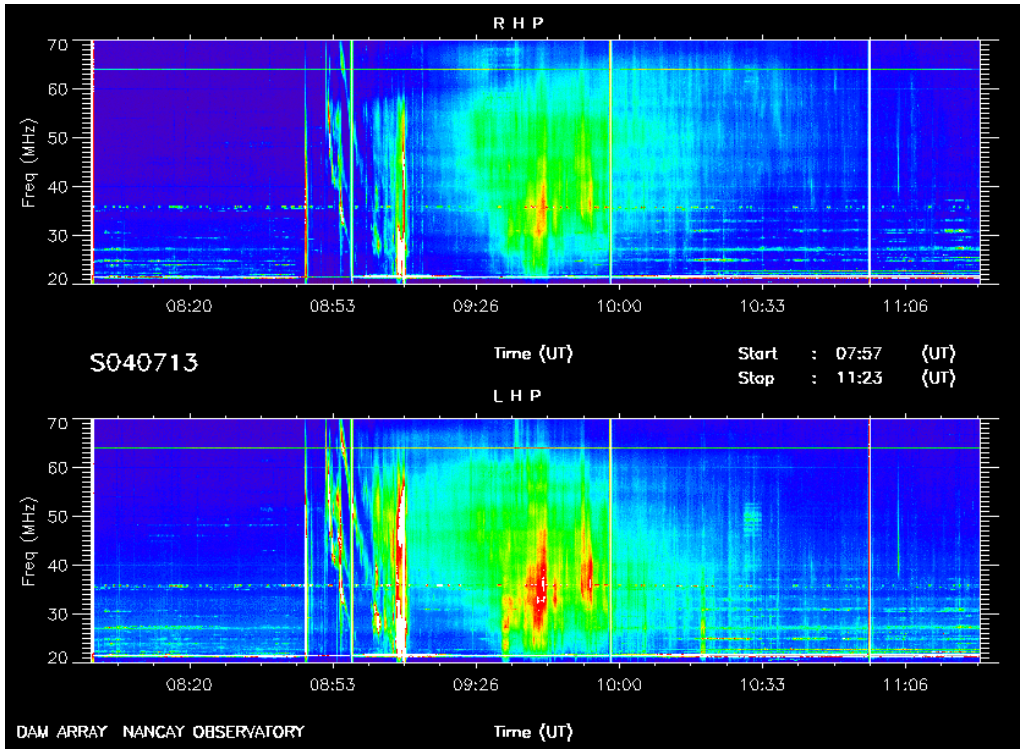

Figure 33. Right and left polarization of July 13, 2004 Type IV burst according to Nancay data.

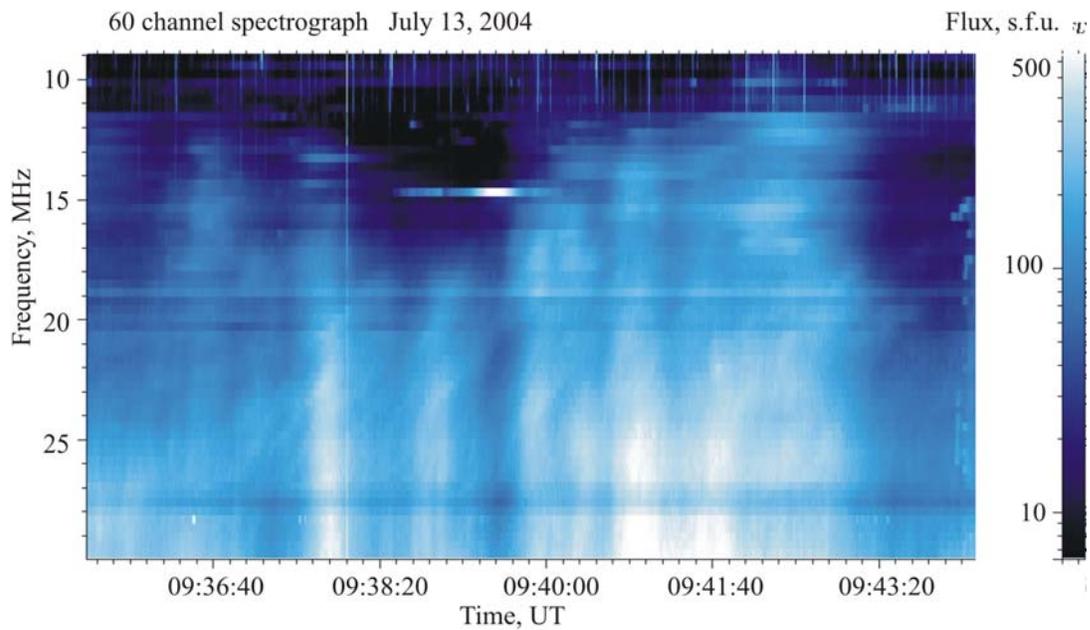

Figure 34. Fiber bursts as a fine structure of July 13, 2004 Type IV burst.

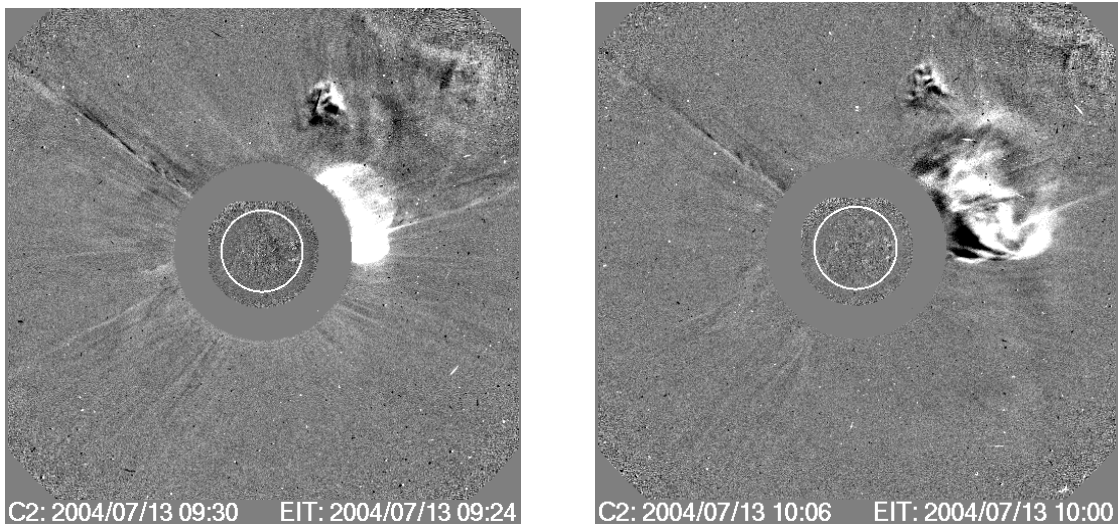

Figure 35. Evolution of a CME associated with the July 13, 2004 Type IV burst observed by SOHO.

*On July 21, 2004*

Data obtained by SOHO shows the CME, which began at 06:24 UT (Fig. 36). This CME is rather slow with a velocity of 419km/s. Judging from the kinetic energy estimations this CME appears to be the weakest of all here discussed events. As in the previous cases, we observe only the low-frequency continuation of the whole Type IV burst (Fig. 37 and Fig. 38). Moreover, in the last case it is already the ending phase of the burst. Nevertheless the flux is very high, reaching the value about 1000s.f.u. At frequency 25MHz it consists of two, and probably even three components separated from each other by 1-2 minutes. The burst terminates at frequency 15-17MHz. At lower frequencies the emission is practically constant and equals 80s.f.u., which is slightly higher than the level of the quiet Sun. On the background of the Type IV burst the lane of Type II burst is seen (Fig. 37). Its drift rate is near 10kHz/s. Judging from NDA data, the drift rate of the Type IV burst is the same. During these bursts a number of Type III-like bursts is observed (Fig.39). They have drift rates from 7 to 12MHz/s and durations 1-2s that is usual for decameter wavelengths. After the Type II bursts there were further observed drift pairs, S-bursts and then spikes. The fast drifting bursts, apparently, are connected with the Type II burst. In addition the Nancay data show a high polarization degree.

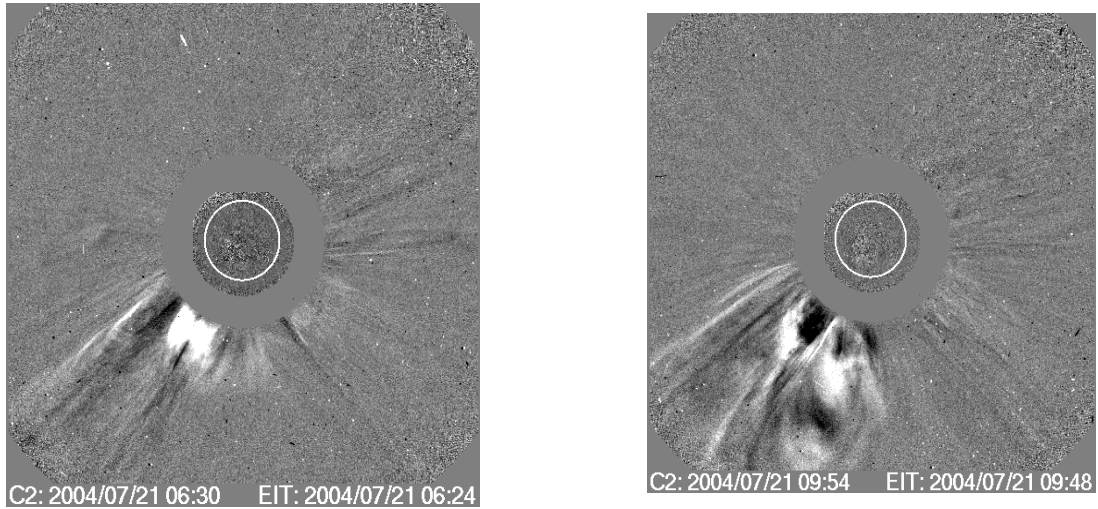

Figure 36. The CME (SOHO data), which initiates Type IV burst on July 21, 2004.

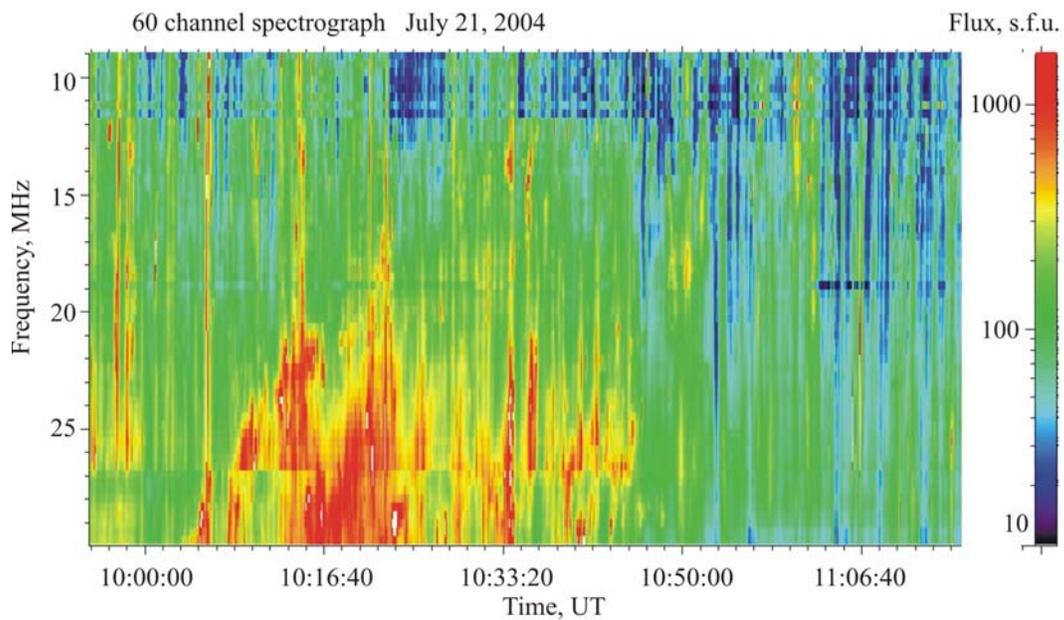

Figure 37. Decameter Type IV bursts on July 21, 2004 observed with UTR-2. Three harmonics of Type II burst are seen at frequencies 20-25MHz.

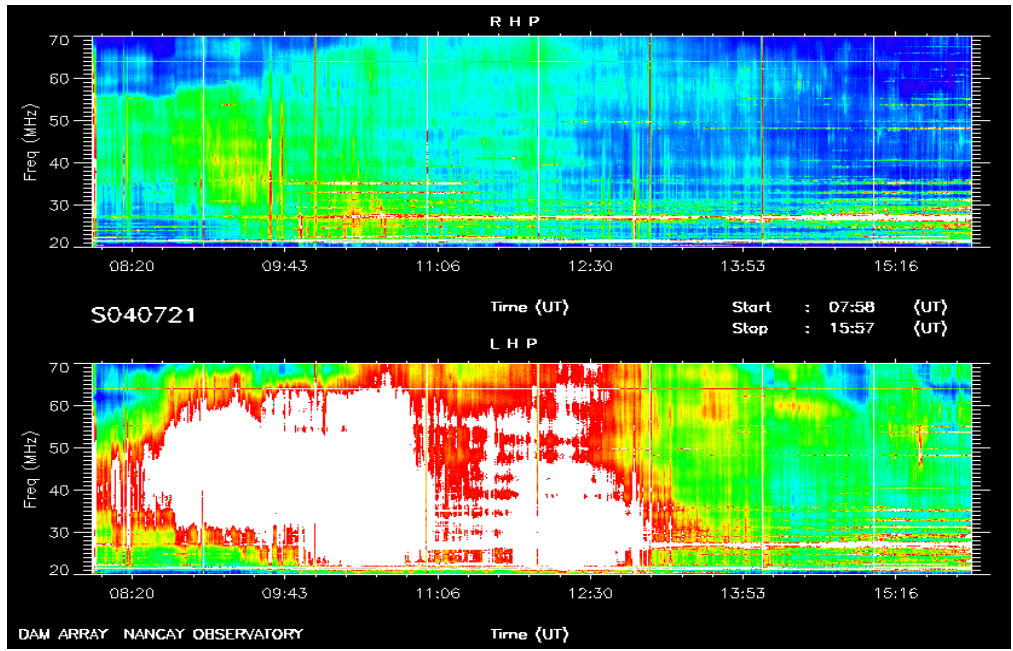

Figure 38. Type IV bursts observed on July, 21, 2004 at higher frequencies (NDA data)

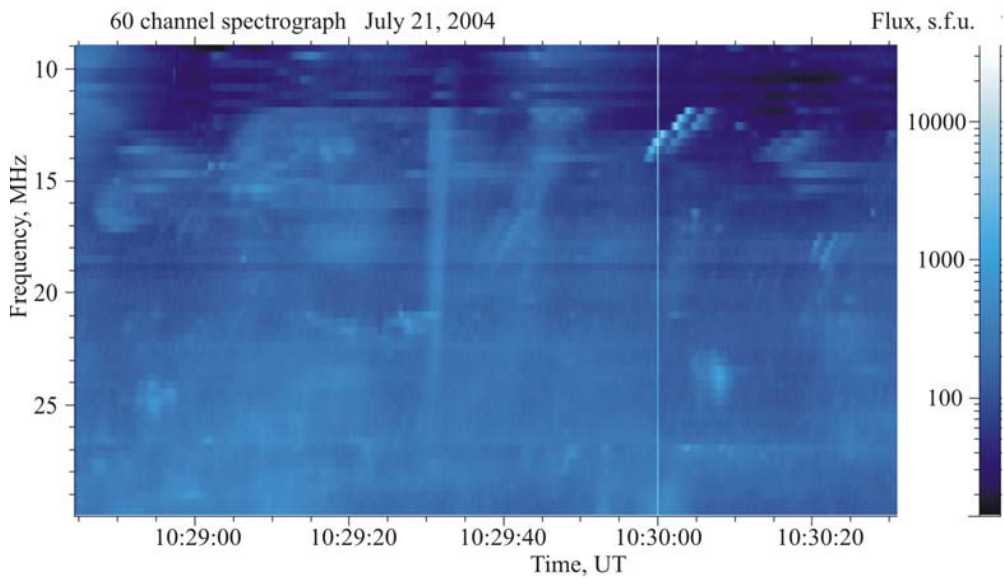

Figure 39. Fine structure of Type IV burst in the form of Type III-like bursts (example at 10:29:30) and drift pairs.

*On July 22, 2004*

An extremely interesting event (Fig. 40) was observed over the frequency range 10-30MHz on July 22, 2004. The increased radiation (with fluxes of more than 100s.f.u.) already existed by the start time of observations. Apparently, the Type IV burst has begun shortly before 08:20 UT. This assumption is supported by data from SOHO (the CME with the speed of 900km/s is visible at the southern part of the solar disk at 08:24 UT) (Fig. 41) and also by NDA data (Fig. 42).

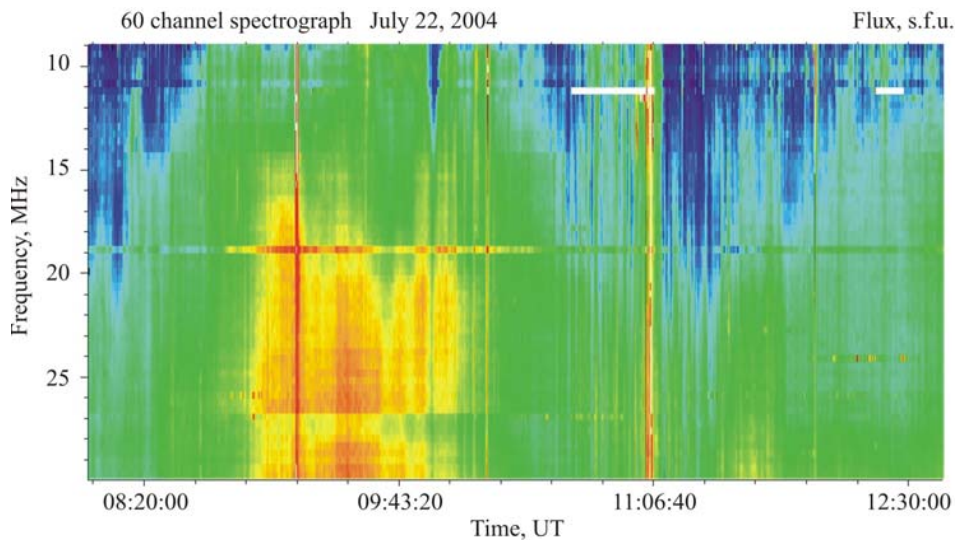

Figure 40. Decameter Type IV burst (UTR-2 data).

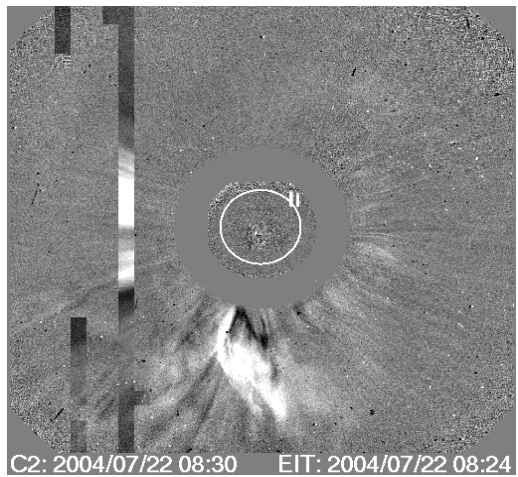

Figure 41. CME, associated with Type IV burst.

At frequencies 10-30MHz the emission intensity is smoothly rising from 08:20 UT till 09:30 UT reaching the maximum value of about 2000s.f.u. (Fig.43). After the maximum point there is an intensity decrease lasting approximately till 10.30 UT. Then one can see two weaker (400 and 500s.f.u.)and shorter (45 and 30min respectively) intensity "hills". By the end of our observations on this day the Type IV bursts still exists (according to the NDA data it lasts till 16:00 UT). At the time of the main maximum of the burst intensity the instant spectrum shows a steep decrease in flux from 2000s.f.u. at 27MHz down to 100s.f.u. at 10MHz (Fig. 44). Nearby the peak intensity point the emission is practically continuous with no fiber bursts visible, although their tracks can be seen. Far aside from a maximum fiber bursts can be detected more reliable. In this case their drift rates are in the majority -2MHz/s. Sometimes there are observed fast drifting sub-bursts, which can be attributed to the class of Type III-like bursts [22]. And, if Type III-like bursts during a storm have durations 1-2s, these bursts inside the Type IV burst have significantly greater durations - up to 10-15s. After 10.00 UT the fiber bursts in emission and absorption and with positive and negative drift rates are distinctly observed. The fiber bursts have a duration of about 10-20s.

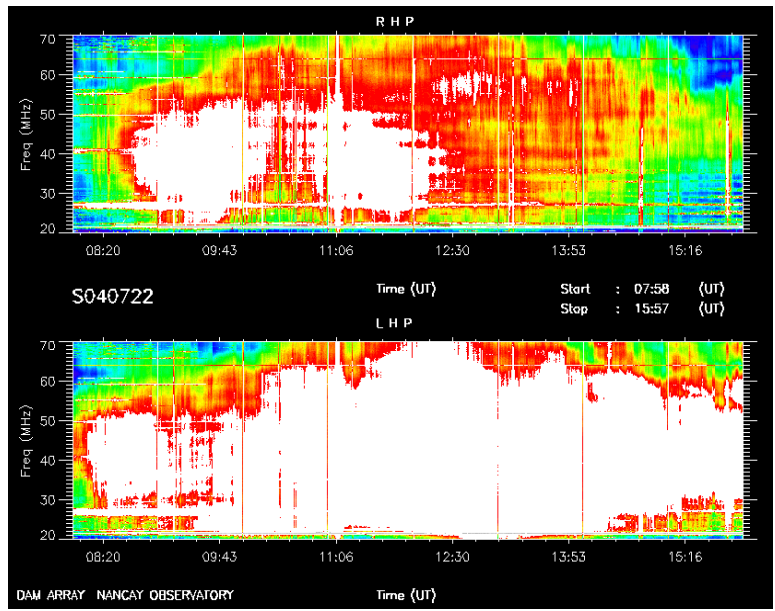

Figure 42. Type IV burst at frequencies 20-70MHz (NDA data).

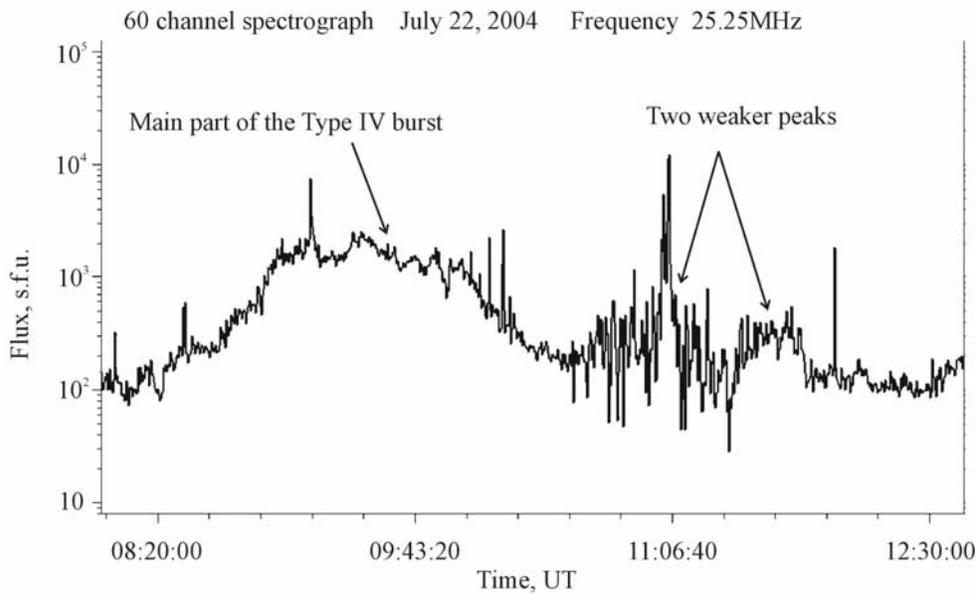

Figure 43. The time profile of Type IV burst with two increased regions (at 11:05 UT and 11:40 UT).

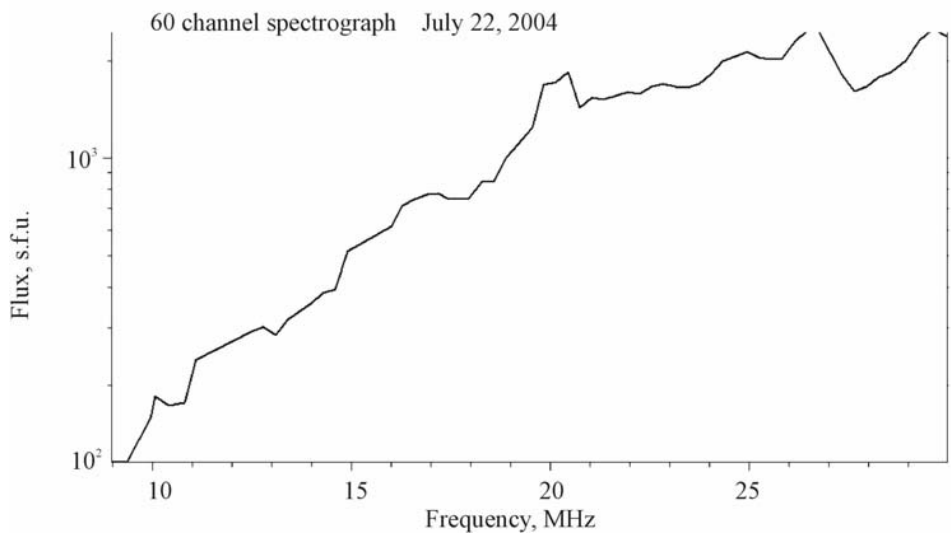

Figure 44. Frequency profile of radio emission at maximum phase of Type IV burst.

An interesting burst is observed at 09:53:53 UT (Fig. 45). This is an absorption burst simultaneously visible in the whole frequency band from 10 to 30MHz. At higher frequencies its duration is 1min. 20s, and at lower frequencies (9MHz), it becomes equal to 6min 20s. While the continuum emission is in the order of 1000s.f.u., the flux of absorption burst falls from 800s.f.u. at 25MHz down to 10s.f.u. at 9MHz.

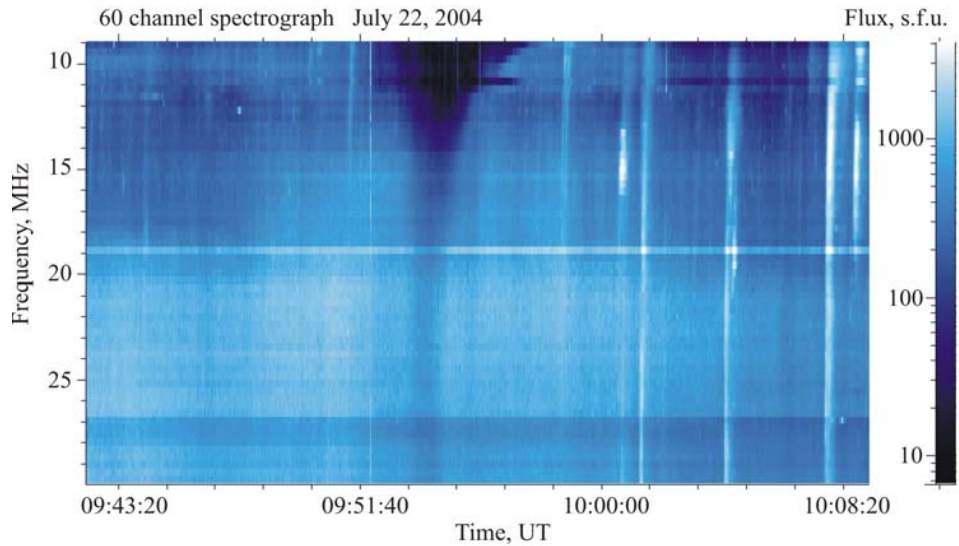

Figure 45. Burst in absorption against continuum radio emission of Type IV burst.

An extremely interesting property of this Type IV burst is a highly populated zebra-structure (unlike the higher-frequency zebra-structures in our case it is similar to finger-prints) (Fig.46). It begins at 08:10 UT and lasts with interruptions till 11:40 UT. The zebra-structures are observed in groups of 5-6 elements and more. The maximum number of elements in a group reaches 38. The stripes of zebra-structure are observed in a frequency range of 14-30MHz and may occur both in emission and absorption. For separate stripes the frequency band makes nearby 5MHz, and in rare cases it can reach up to 10MHz. Drift rates of zebra stripes can be both positive and negative, and sometimes infinite (Fig. 46). There are cases when the same stripe at higher frequency has negative drift rate, and at lower frequency positive one. The absolute average values of zebra structure drift rates are 0.5MHz/s. The distance between stripes in emission (or absorption) may range from 1.2s up to 1.3s in the beginning of the burst and remains practically constant for the given group. For the group of zebra stripes the distance between maxima is 2s at 09:03 UT, and 2.5s at 11:36 UT. Our records distinctly show, that stripes in emission are alternating with stripes in absorption (Fig. 47). The flux during emission phase equals 284s.f.u and during absorption falls down to 245s.f.u. Thus the mean flux of the Type IV bursts during this stripes group (08:38 UT) makes 252s.f.u. We must note that the fluxes in emission and in absorption for a given group remain approximately constant. The same properties are observed also in other groups of zebra-structures of this Type IV burst.

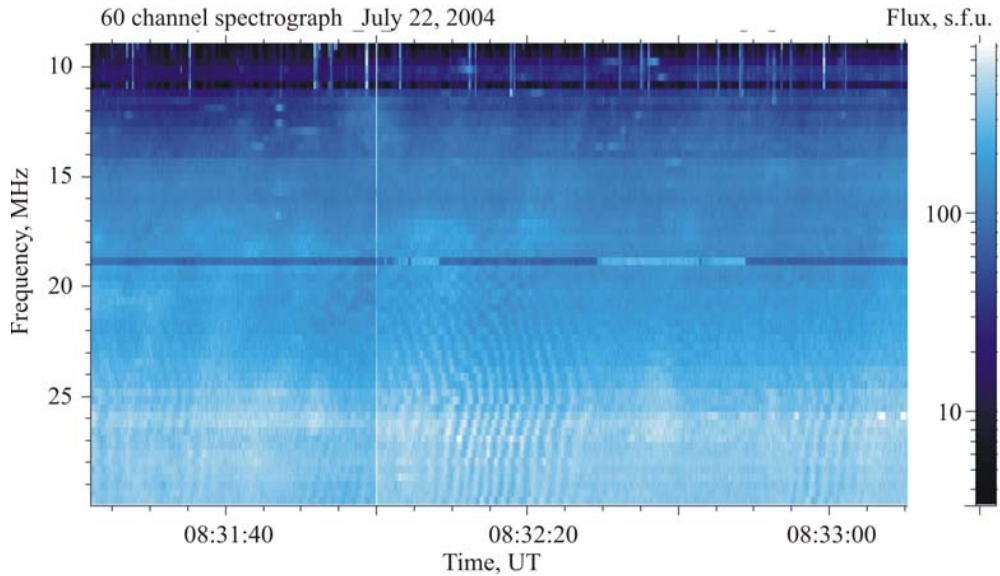

Figure 46. Three groups of zebra patterns.

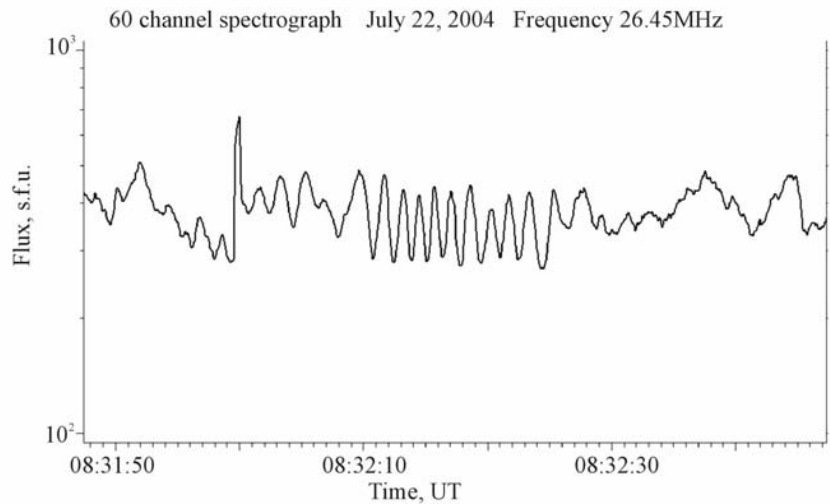

Figure 47. Succession of stripes in zebra structure in emission and in absorption.

*On July 23, 2005*

This Type IV burst (Fig. 48) is interesting first of all by the fact that it doesn't follow the apparently connected with a Type II burst, as it usually happens. The associated Type II bursts are observed against a background of the Type IV burst. In other words, two bursts are superimposed. Also unlike the previous cases there is no group of intensive Type III bursts before the Type IV burst. Taking into account low flux and short duration of both bursts it may be supposed that the initial event was also not so powerful. Data from SOHO (Fig. 49) show that both the bursts, apparently, are connected with the CME, which took place at 09:30 UT and had a velocity of 671km/s. The Type II burst drift rate is 20-30kHz/s. It consists of several lanes. Also this Type II burst has fine structure in the form of sub-bursts with durations 1-2s and drift rates 4-6MHz/s (sometimes sub-bursts with infinite drift rates can be detected). The Type IV burst is most likely a moving burst. It propagates as a whole from high towards low frequencies with a drift rate of 10kHz/s.

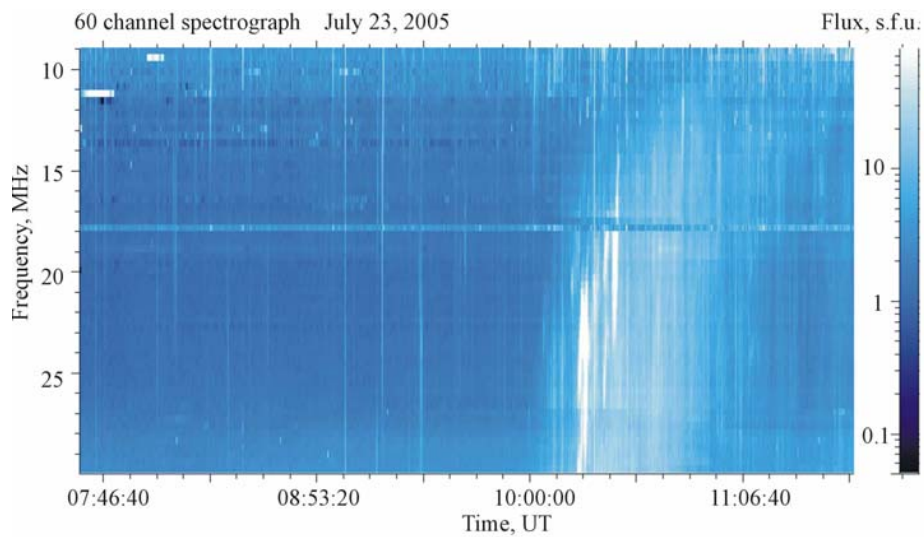

Figure 48. Decameter Type II-IV burst.

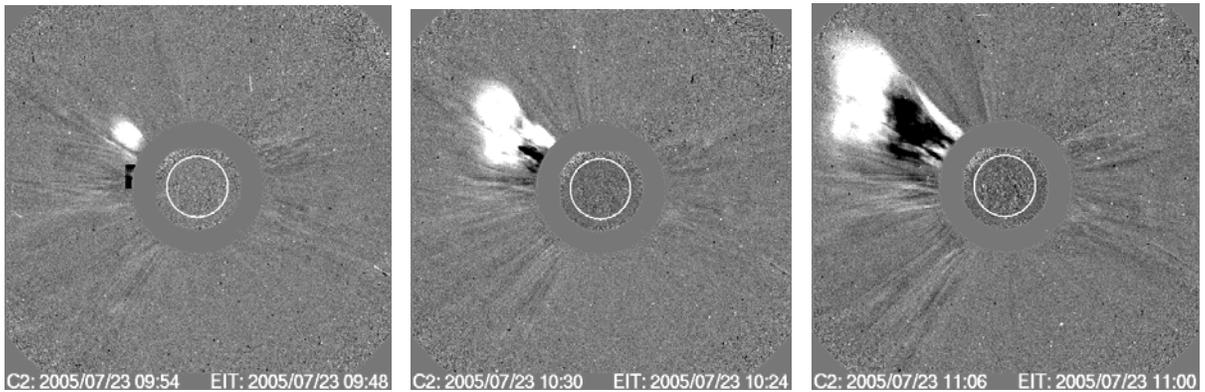

Figure 49. CME activity on July 23, 2005.

In addition to the main part (maximum flux is 20s.f.u. at 25MHz) there can be distinguished also a less intensive part (maximum flux is 3s.f.u.) (Fig.50), which is observed at 11:40 UT and exhibits practically no drift. The time profiles of both parts of the burst are symmetrical, and while the first part has a duration of about 1 hour, the second part lasts only 20min. The slopes of instant spectra of two parts of the bursts are also different. In the first case the flux decreases (approximately by an order of magnitude) with decreasing frequency, while in the second case the flux remains practically constant in the whole frequency band. The Type IV burst is limited by a "bottom" frequency 11MHz, and splits in two lanes at 14MHz. The main part of the Type IV burst consists of numerous faint fiber-bursts (Fig. 51).

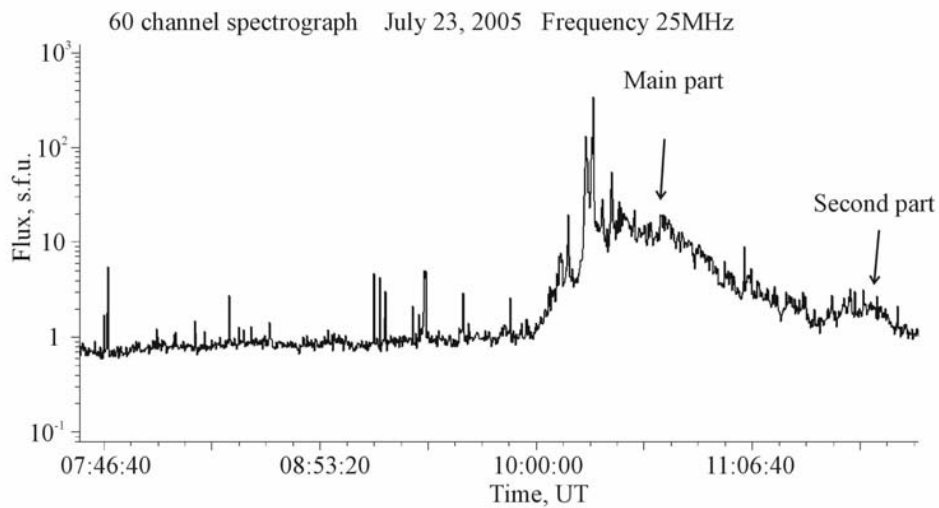

Figure 50. Type IV burst profile with two enhanced regions.

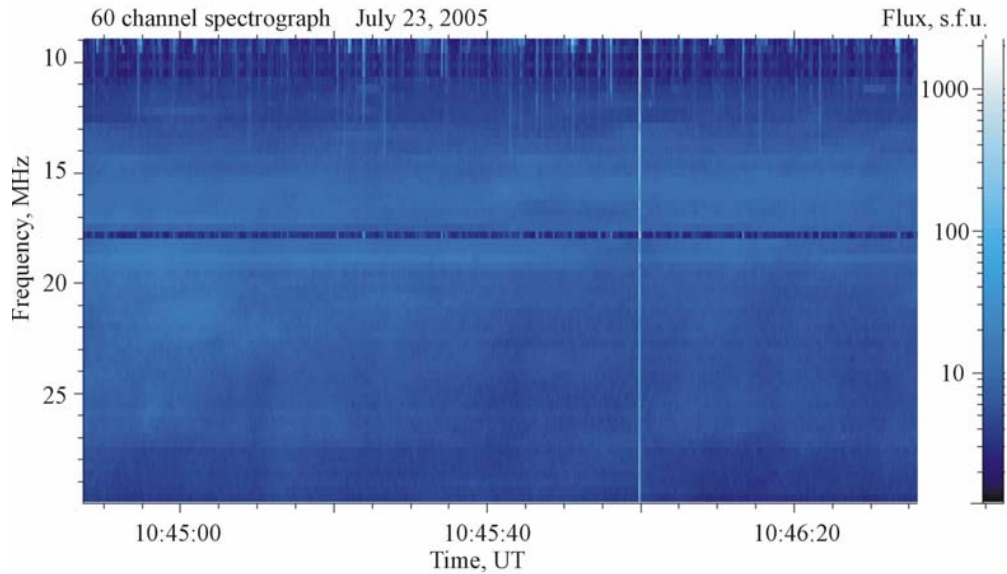

Figure 51. Fragment of continuum radio emission of Type IV burst.

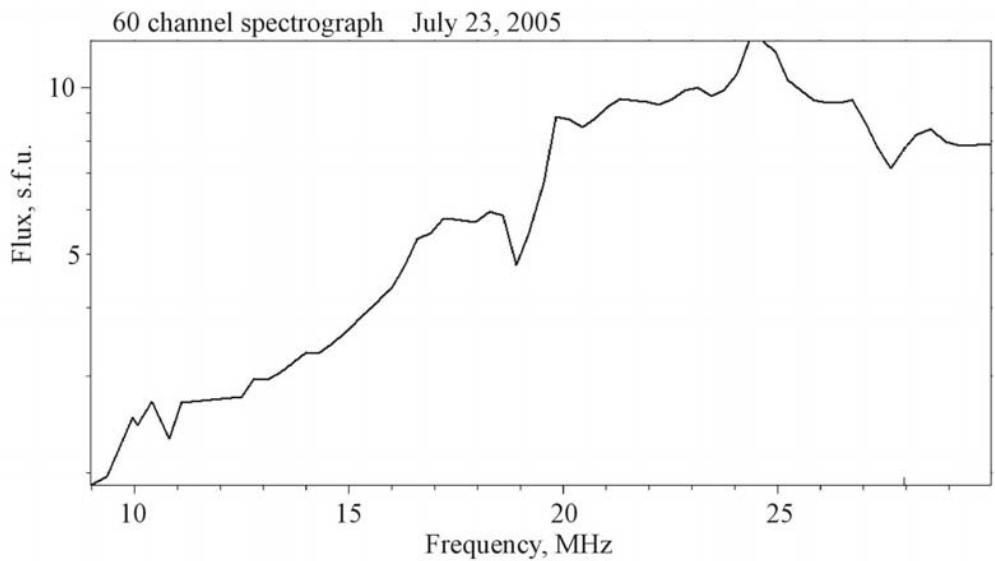

Figure 52. Frequency profile at maximum phase of Type IV burst.

(The flux difference between the fiber burst maxima and nearby minima is factor 1.5 ). Their durations are about 10s, drift rates 1-2MHz/s and the flux at its maximum is 10s.f.u. It is interesting, that at its maximum the flux remains almost constant at frequencies from 30 down to 20MHz and then steeply falls towards lower frequencies (Fig. 52). On the contrary, at the end of our observations on that particular day July 23, 2005, the flux is practically constant at low frequencies and falls off towards higher frequencies.

*On July 27, 2005*

The event July 27, 2005 (Fig. 53) started at 07:46 UT. As many other Type II and Type IV bursts it follows a group of Type III bursts having durations of about 10-20s and a peak flux up to 700s.f.u. On the background of these Type III bursts one can see a plenty of fiber bursts (Fig. 54) with durations up to 10s and drift rates 10-20kHz/s. This drift rate is inherent to the Type II bursts at the decameter wavelengths though apparently there were no Type II bursts visible on that day.

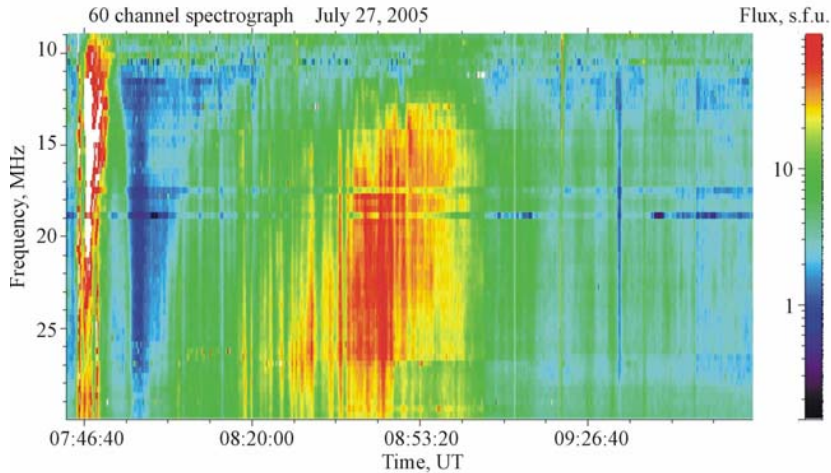

Figure 53. Type IV burst on July 27, 2005.

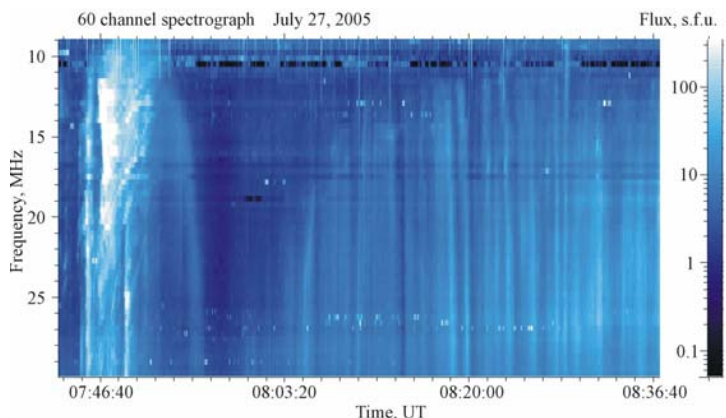

Figure 54. The beginning Type IV burst followed by a group of Type III bursts, fiber bursts and absorption region.

There are no tracks of their presence also at other frequencies, though a CME with velocity 533km/s was detected by SOHO at 07:54 UT (Fig. 55). After the group of Type III bursts and fiber bursts the intensity of radiation decreases and a shading area appears. Over the whole frequency band from 10 to 30MHz the flux in this area is constant and equal to 0.9s.f.u. The width of the shading area at 30MHz comprises 2min 30s, and at 13MHz already 7min 30s. After the shading area at 08:00 UT the Type IV emission starts rising and lasts till 08:50 UT (Fig. 56). At this time the flux has its peak value of 60s.f.u, which is practically constant over the whole bandwidth of the burst. On a falling branch of the Type IV bursts, which lasts till 11:00 UT, it is possible to distinguish some large-scale local increases in radiation with durations of about 30min.

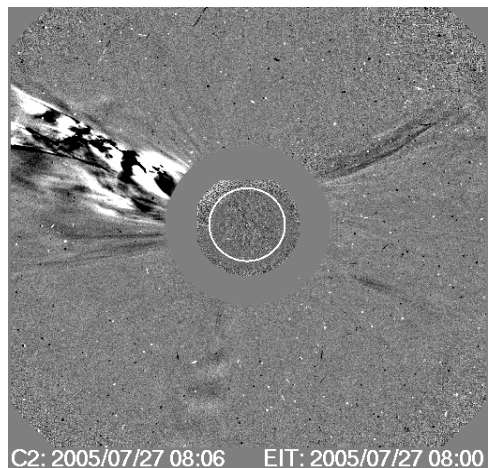

Figure 55. CME associated with July 27, 2005 event (SOHO data).

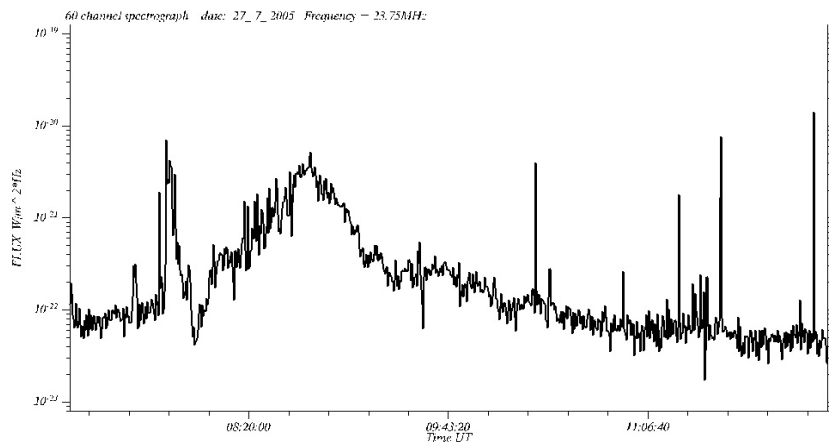

Figure 56. Time profile of the Type IV burst.

More fine structure of the Type IV burst represents faint fiber bursts with durations from 10 to 50s and both positive and negative drift rates. Moreover, a considerable number of such bursts in absorption is observed including those with the positive drift rate (so the burst at 08:14 UT had a drift rate of +8.58MHz/s and duration around 1min). On a descending branch of the Type IV burst there can be seen more contrast sub-bursts in emission and absorption with reverse drift rate (from low to high frequency). A corresponding example of such bursts is shown in Fig. 57. The first sub-burst (about 09:31 UT) has a drift rate of +1.2 MHz/s and duration 13s. This sub-burst has the knee-bend at frequency 15-16MHz and at low frequencies it has already a forward drift rate -0.36MHz/s. The drift rate of the subsequent burst in absorption is 3.0MHz/s and it lasts about 10s. The next sub-burst has a drift rate of 1.62MHz/s and a duration of 14s. And finally the last burst in absorption drifts with 9.47MHz/s and has a duration changing almost linearly from 40s at high frequencies to about 1min 30s at low frequencies. The time profile (Fig.58) is nearly symmetrical although there is a tendency of a more flat descending branch.

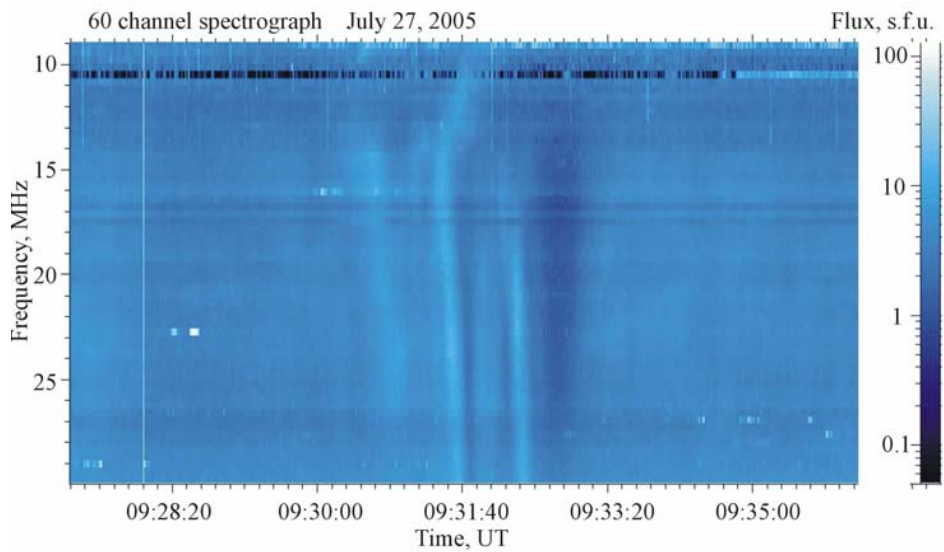

Figure 57. Bursts in emission and absorption with positive drift rates.

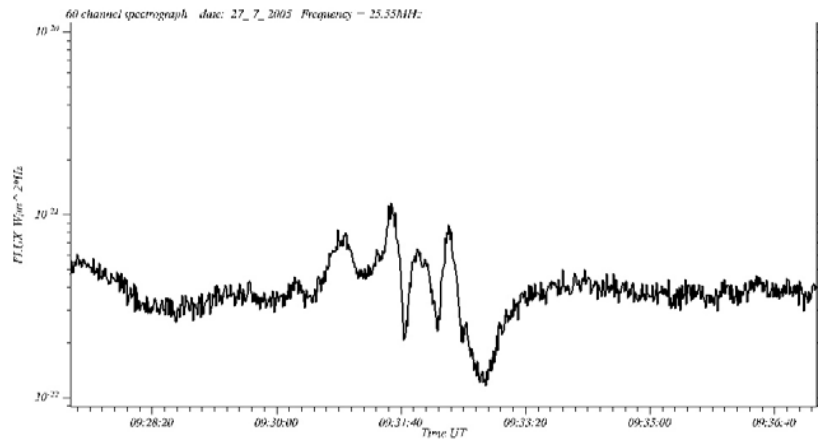

Figure 58. Time profiles of bursts in emission and in absorption.

*On July 28, 2005*

This Type IV burst (Fig. 59) is characterized by the absence of any associated Type III and Type II bursts. Nevertheless SOHO shows the CME started a little bit earlier, at 07:00 UT (Fig. 60) with a velocity of 573km/s. The front edge of this burst drifts with the rate close to 10kHz/s from high to low frequencies. The lowest frequency, reached by the Type IV burst is - 12-13MHz.

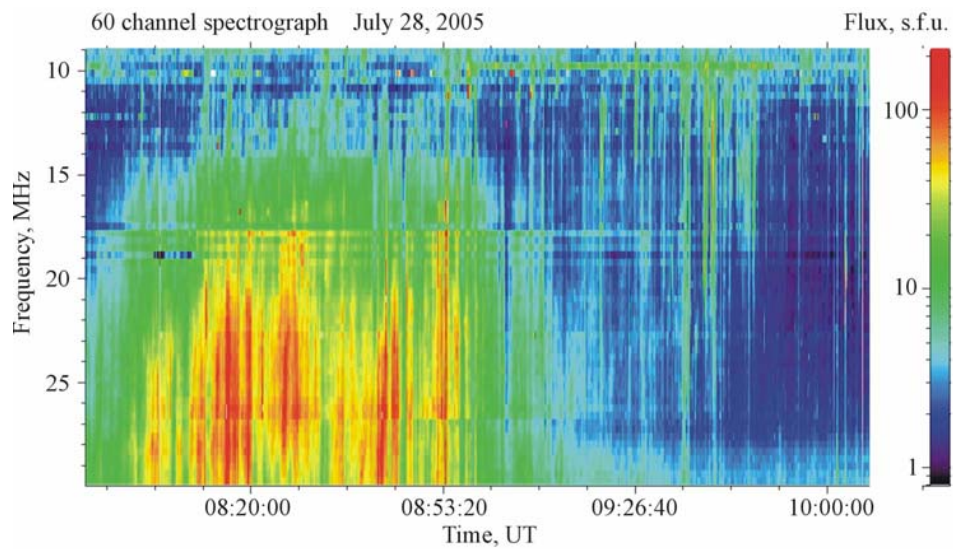

Figure 59. Type IV burst on July 28, 2005.

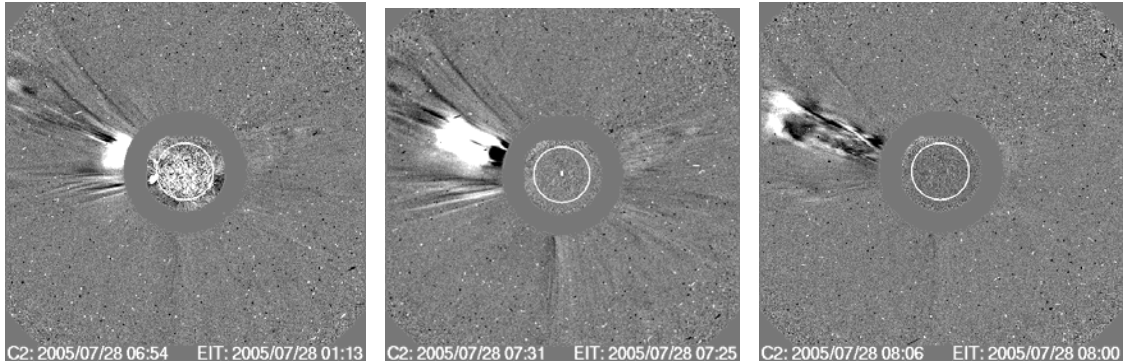

Figure  60. Evolution of CME from SOHO.

At frequencies above 18-19 MHz the burst consists of two practically identical parts separated in time from each other by approximately 30min (Fig. 61). The instant spectrum at  maximum flux point  shows almost a constant flux from 30 down to 23MHz, which is equal to 200s.f.u. (Fig. 62). And from 23 to 13MHz the flux steeply falls down to 6s.f.u. Like the previous Type IV burst (on July 27, 2005) this burst has a fine structure in the form of fiber bursts, however they are less distinct. The  duration of these fiber bursts is 5-10s, and their drift rates in the majority are typical for common Type III bursts in the decameter wavelengths, i.e. 2-4MHz/s. At the same time sub-bursts with the reverse drift rates are sometimes observed (Fig.63).

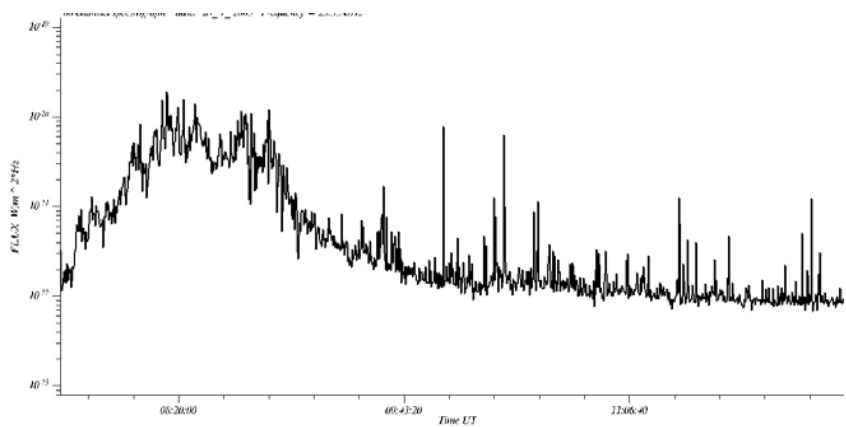

Figure 61. Time profile of Type IV burst observed on July 28, 2005 at 25MHz.

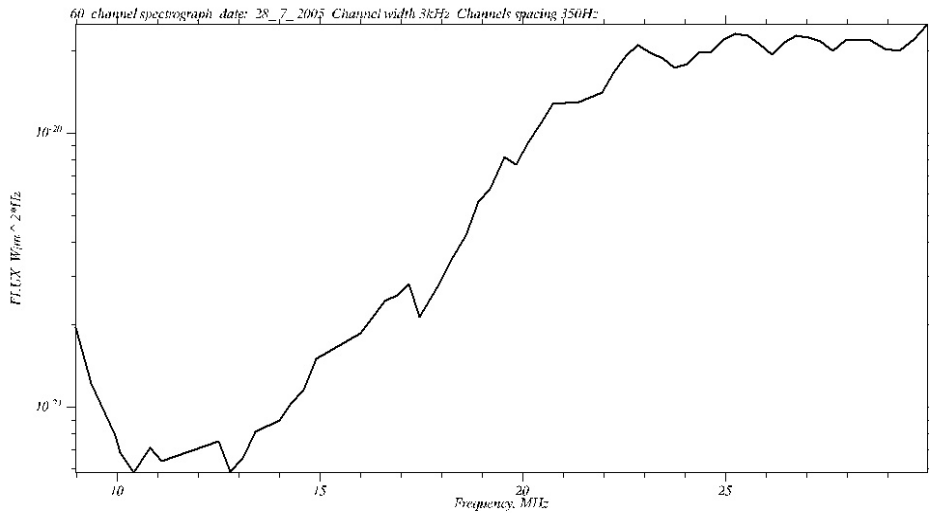

Figure 62. Frequency profile in the maximum phase of Type IV burst.

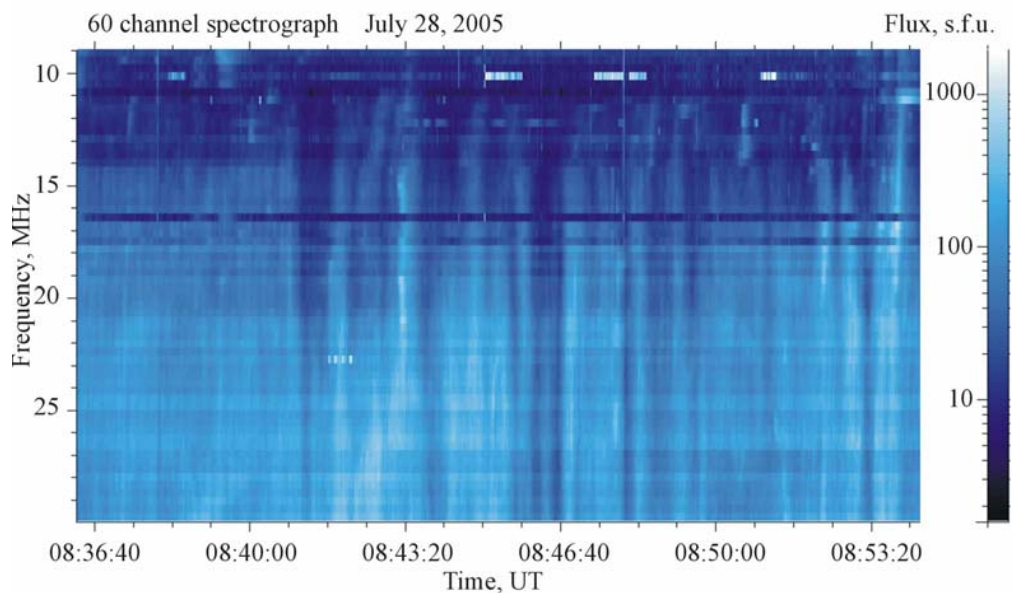

Figure 63. Fine structure of Type IV burst in the form of fiber bursts with positive and negative drift rates in emission and absorption.

In the majority sub-bursts have a restricted frequency band of existence, given by 3-8MHz. These bursts are more clearly seen on the tail of the Type IV burst, and in addition there are also bursts in absorption with positive and negative drift rates (Fig. 63).

*On July 31, 2005*

This event we also file to the class of Type IV bursts since an increase of continuous emission is observed from the quiet Sun background up to a maximum value of 20 s.f.u. at 10:00-10:15 UT (Fig. 64). After that the flux decreases. The total estimated duration of the Type IV burst comprises not less than 7 hours. At the peak phase of the burst the flux practically does not vary with frequency. The prominent feature of this burst is that it consists of both the fiber bursts, and a number of forward and reverse drift pair bursts. Fiber bursts have various drift rates (Fig. 65) (positive, negative and infinite), durations (from 2 s up to 10-20s), and frequency bands (from a few MHz up to 20MHz). In some cases drift pairs appear to be brighter with fluxes several times the continuous emission flux. This Type IV burst is associated with a CME (Fig. 66), which was registered at 06:30 UT and is also connected with the active region in the eastern part of the solar disk.

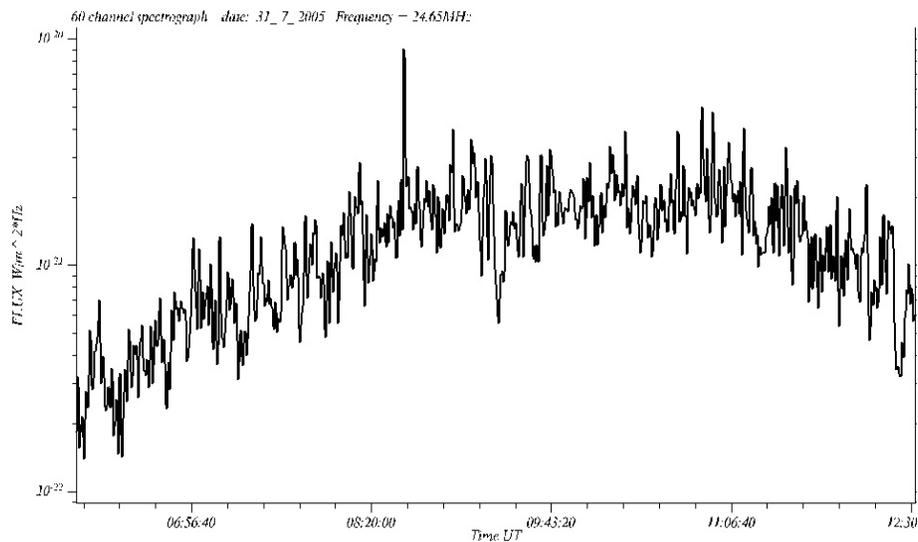

Figure 64. Time profile of prolonged Type IV burst.

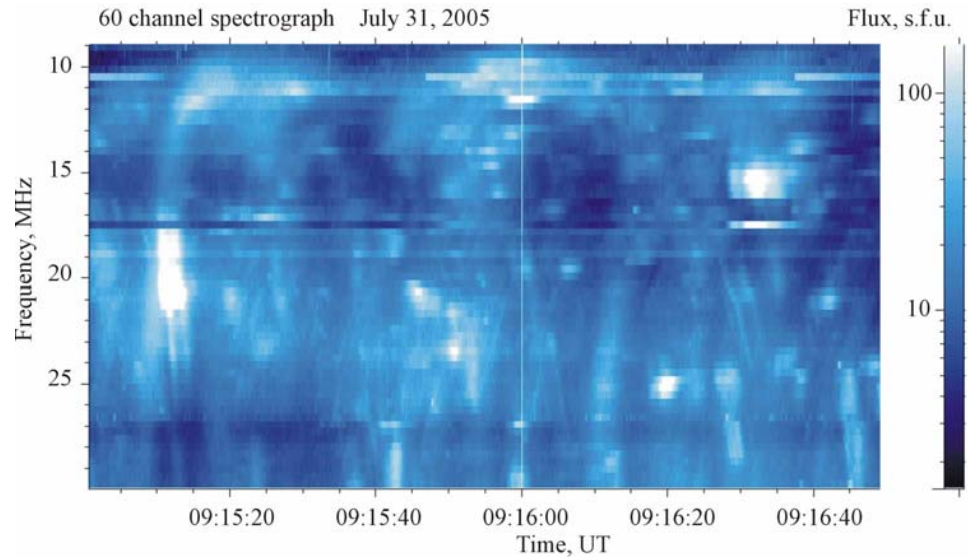

Figure 65. Fine structure of Type IV bursts in the form of fiber bursts and drift pairs.

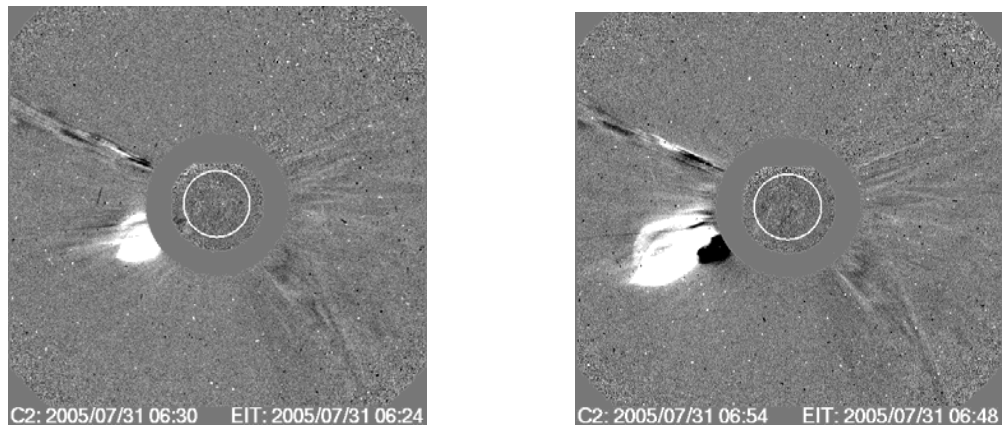

Figure 66. Dynamics of CME at 06.24 UT on July 31, 2005.

*On July 6, 2006*

This event (Fig. 67), apparently, is initiated by CMEs, which took place in series at 06:54 UT and 08:54 UT (Fig. 68). The group of Type IIIb and III bursts and subsequent Type II burst are connected with the first CME.

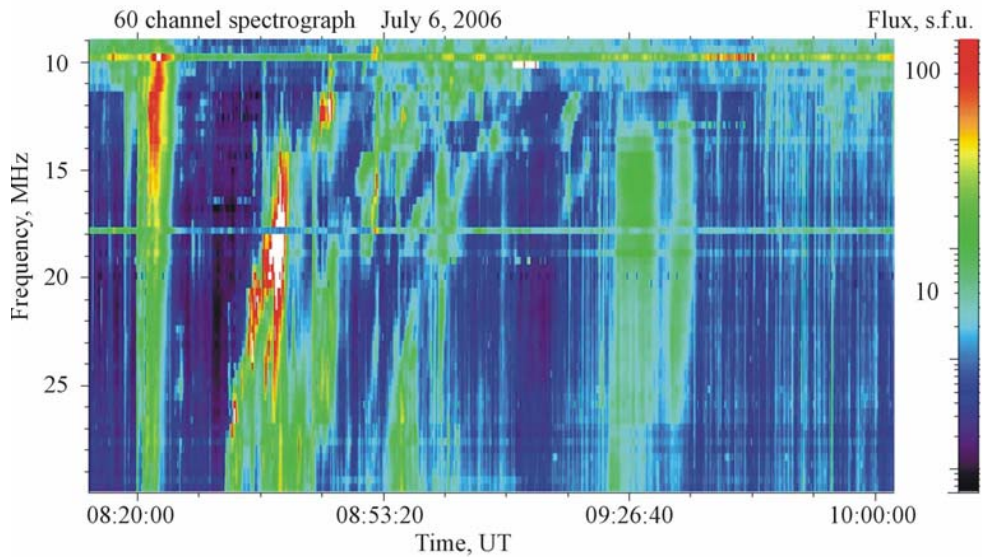

Figure 67. Type IV burst on July 6, 2006.

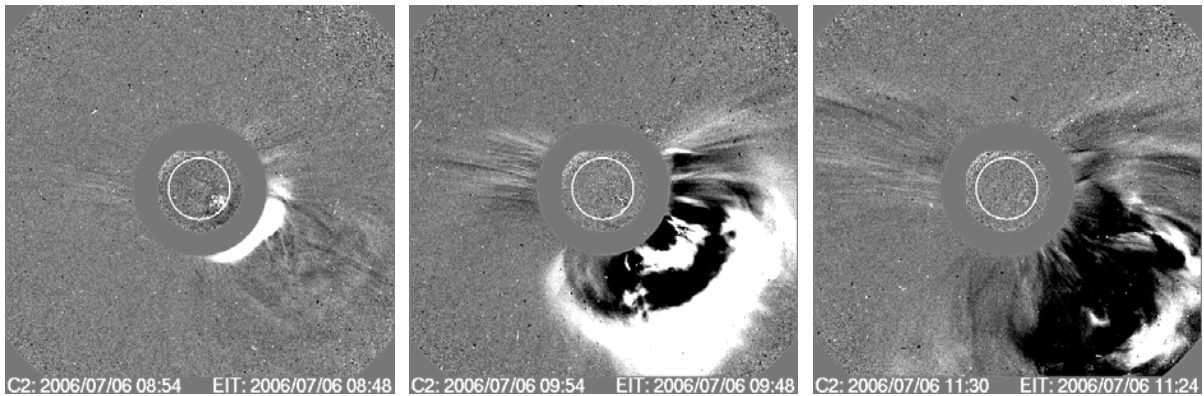

Figure 68. CME associated with July 6 Type IV burst.

The Type III and IIIb bursts (at 08:20 UT) are very intensive with fluxes up to 1000 s.f.u. (Fig.69). The total duration of the group is about 2min 30s, and drift rates are 2-4MHz/s. After this group the emission level drops down to a level practically equal to the quiet Sun background (1 s.f.u.). Then, at 08:31:40 UT the Type II burst starts at 30MHz (Fig. 70). A short time later its second harmonic appears. The drift rates of both harmonics are 20-30kHz/s. Sub-bursts, which form both the first and second harmonics, have various drift rates, ranging from - 0.5  to -3MHz/s. Their durations are 2-5s, and sometimes more.

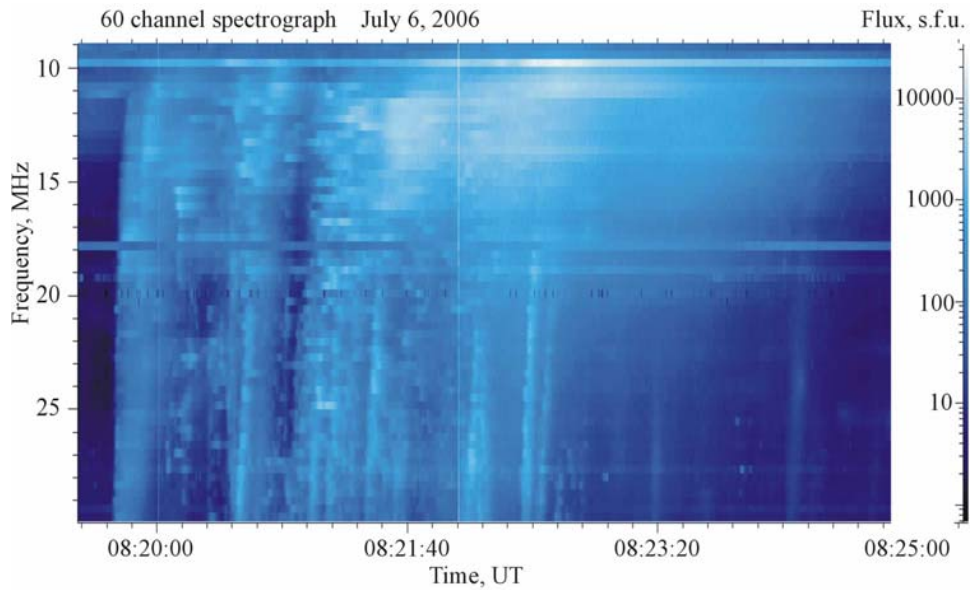

Figure 69. Group of Type III bursts as a precursor of Type IV burst.

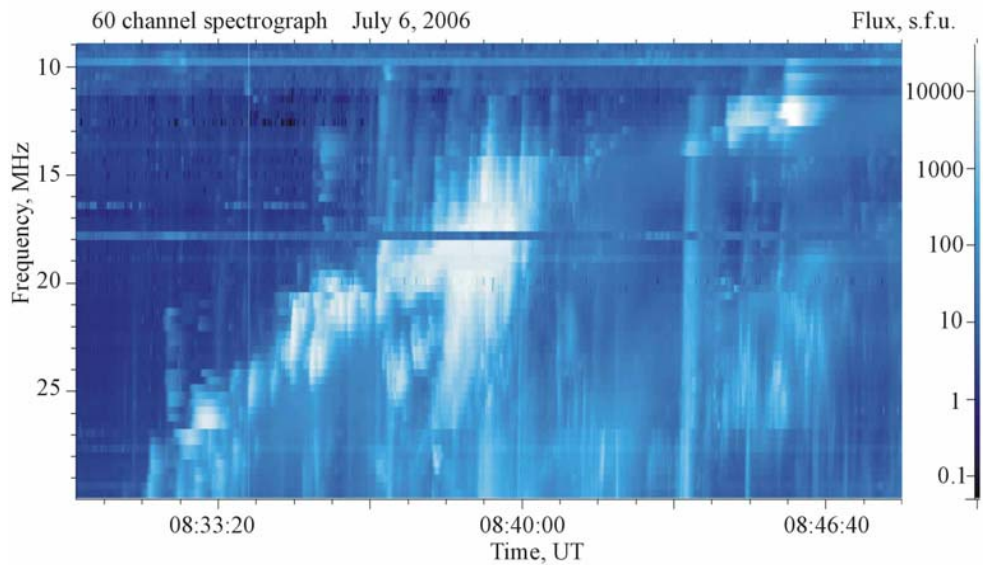

Figure 70. Fundamental and harmonic emission of Type II burst before Type IV burst.

Right after this an increased level of emission is observed (Fig. 71). It consists of 5-6 long-lived broadband flux increases reaching sometimes 100s.f.u. and spaced in time by approximately 30min. In general these increases do not consist of very intensive fiber bursts, which in the majority have durations 3-10s and drift rates of 1-3MHz/s, which is slightly less with regard to common Type III bursts. We also observed some sub-bursts with drift rates of 6-10MHz/s and durations of 1-2s. They resemble fast Type III-like bursts, which usually occur during common Type III bursts storms when the associated active region is close to the central meridian [22]. Among all mentioned 6 flux increases two cases (with maxima at 09:27:30 UT and 09:34:10UT) are of special interest (Fig. 72). Contrary to the other maxima, which consist of bursts resembling common Type III bursts, these two emission "hills" represent continuum emission with fluxes about 100s.f.u. The first one has a total duration of 7 min and splits in two elements above 18MHz. The second one has a duration near 3min. Both features drift with rates of 0.3MHz/s.

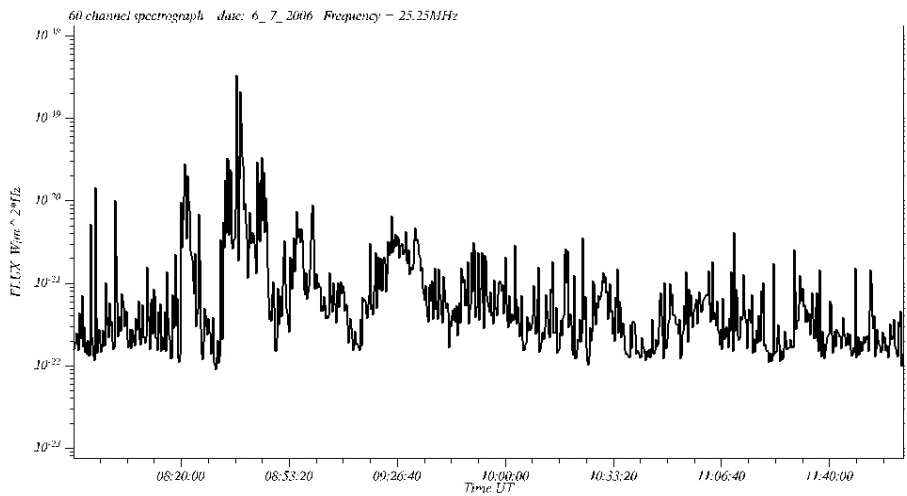

Figure 71. Time profile of Type IV burst.

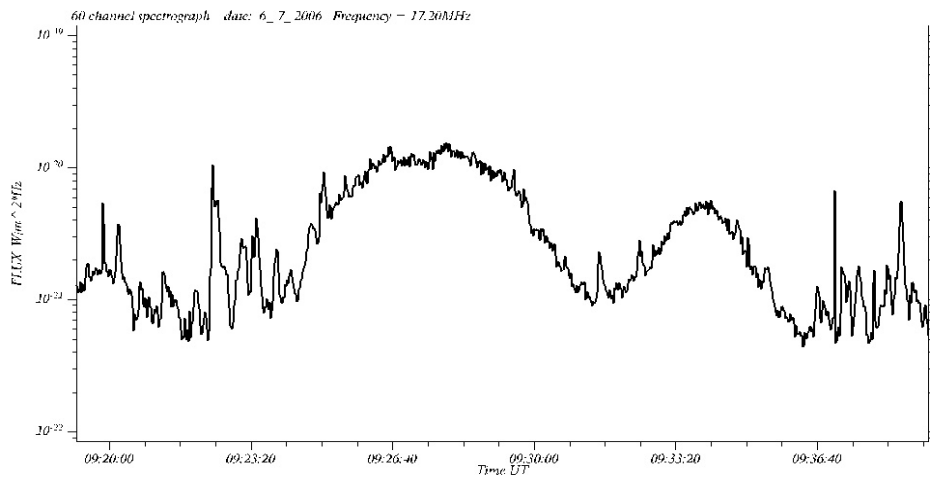

Figure 72. Two continuum "hills" in Type IV burst.

## Discussion

All discussed events in this chapter can be categorized to the class of Type IV bursts due to the following reasons:

- these bursts have a noticeable continuum component, which, as a rule, rises at the beginning of the bursts and falls at the end of it;
- in the majority of cases these bursts follow the Type II bursts or the latter are seen on the background of the first;
- in all cases these bursts are associated with CMEs.

*The succession of the events accompanying the Type IV bursts in the decameter wavelengths.*

Preceding a Type IV burst by 10 – 30 minutes, a group of very intensive Type III bursts is observed. For decameter wavelength band they have standard durations and frequency drift rates. Just at the end of such a group and/or on its background, fiber bursts are seen with drift rates 20-30kHz/s and durations of 3-5s. (Mentioned drift rates are inherent to Type II bursts at the decameter waves). The group of Type III bursts and fiber bursts are followed by a shading area with a duration of some minutes, which uses to increases in several times towards the lower frequencies.[h3] Their drift rates can be either positive or negative and the fluxes at the minimum correspond to the background, as observed prior to the whole event. Just after the bursts shading area a Type II burst begins. The Type II bursts usually consist of fundamental, second and sometimes third harmonic emissions. A Type IV burst starts just after Type II burst or slightly earlier.

*Properties of the decametric Type IV bursts.*

- durations of Type IV bursts at decameter wavelengths vary from some tens of minutes to some hours;

- moving as well as stationary Type IV bursts are observed. In the case of moving ones their drift rates are close to the drift rates of the preceding and/or accompanying Type II bursts;

- as a rule, Type IV burst are formed by two or more components with the first component been the most intensive. Sometimes the sequence of components resembles an oscillation process;

- radio emission fluxes of Type IV bursts at maximum reach 2000s.f.u., however there are also weak Type IV bursts with fluxes of several tens of s.f.u. only;

- the Type IV burst spectrum taken at the time of maximum intensity shows that the flux can either fall or rise towards lower frequencies. Sometimes the flux remains constant over the whole frequency range from 10 to 30MHz. The first case happens when fine structure sub-bursts are more distinctly seen on the continuum background and the drift rates of the sub-bursts are close to those of Type III bursts in storms. In the second case, the less distinct sub-bursts have drift rates smaller than 1MHz/s, i.e. 2-3 times slower than common Type III bursts;

- in rare cases the Type IV burst fine structure represents drift pair bursts, S-bursts, spikes and Type III-like bursts;

- the fine structure known as a zebra-structure is an extremely rare event in the decameter wavelengths range. It is observed during the overall burst extent in groups, with from 5-6 up to 30-40 stripes in each of the groups;

- fiber bursts have different drift rates, both positive and negative. The fiber bursts in absorption are not uncommon;

- large scale absorption bursts are observed against Type IV burst background. These fast drifting bursts have durations of several minutes.

*Fine structure properties of Type IV bursts*

The most frequently seen fine structure of Type IV bursts are fiber bursts, which have parameters comparable to common Type III bursts. In particular, their durations are 5-20s and drift rates from less than 1MHz/s up to 5MHz/s. Contrary to the Type III bursts, the fiber bursts may exist both in emission and absorption, may have both positive and negative drift rates and are observed in limited frequency band (from 5-6MHz up to 20MHz).

At higher frequencies (at meter wavelengths) Type IV bursts appear to have fine structure in the form of so-called zebra-structure. We found only one Type IV burst with zebra-structure (July 22, 2004). The groups of zebra-structure stripes last during the full Type IV burst duration, slowly drifting with time towards lower frequencies. In total more than 30 groups of zebra-structure stripes have been observed. At the beginning of the Type IV burst each stripe of zebra-structure has negative drift rate at higher frequencies and positive rate at lower frequencies. And at the end of the Type IV burst the stripes have mainly positive drift rates. Stripes inside a group are spaced by constant time intervals, which equal from 1.5-1.6s, at the beginning of the Type IV burst, to 2.5s by the end of it. Inside one group the stripes in emission and absorption are alternating. An interesting feature of zebra structures in the decameter band is that the fluxes at maxima of stripes in emission (as well as at minima of stripes in absorption) keep constant along a group.

*Polarization of Type IV bursts*

Because of lack of numerous polarization measurements, we cannot provide any definite conclusions on the polarization of Type IV bursts at the decameter wavelengths. The only fact we can state is that although the polarization is rather high for some Type IV bursts (up to 60%), the polarization is smaller with regard to those observed at higher frequencies (e.g. up to 100% at meter wavelengths). We also do not see any monotonous increase of polarization from the beginning to the end of the Type IV burst, as it was found at higher frequencies. We can rather state that the polarization change in time during Type IV bursts may reflect some changes of inner structures inside the Type IV bursts sources.

## Conclusions

Observations of the decameter Type IV bursts with unique sensitivity, high time and frequency resolutions allowed to define the Type IV bursts properties at lowest frequencies, with regard to ground-based observations. Zebra structure at those lowest frequencies was detected. The properties of zebra-structure at decameter wavelengths and at higher frequencies are found to be different. Large-scale bursts in absorption were observed on the background of Type IV bursts.

The observations showed that Type IV bursts at decameter wavelengths on the one hand are characterized by a great variety of properties, such as duration, flux, large-scale structure, and on the other hand have some similar properties – fine structure consisting mainly of fiber bursts, which resemble common Type III bursts.

Further observations of Type IV bursts, including polarization observations in a wide frequency band will support a deeper understanding of this phenomenon in large and small scales.

## Acknowledgements

The authors are grateful to SOHO, WIND and Nancay teams for operating the instruments and performing the basic data reduction, and especially, for the open data policy. This chapter was partially supported by INTAS grant No. 03-5727.